\documentclass{elsart}

\usepackage{graphicx,amsfonts,amssymb,amsmath,latexsym} 

\newcommand{\cal}{\mathcal}

\newcommand{\Wsla}{W\hspace{-0.88em}/}
\def\be{\begin{equation}}
\def\ee{\end{equation}}
\def\ba{\begin{eqnarray}}
\def\ea{\end{eqnarray}}
\def\gta{\mathrel{\raise.3ex\hbox{$>$\kern-.75em\lower1ex\hbox{$\sim$}}}}
\def\lta{\mathrel{\raise.3ex\hbox{$<$\kern-.75em\lower1ex\hbox{$\sim$}}}}

\def\simgt{\mathrel{\raise.3ex\hbox{$>$\kern-.75em\lower1ex\hbox{$\sim$}}}}
\def\simlt{\mathrel{\raise.3ex\hbox{$<$\kern-.75em\lower1ex\hbox{$\sim$}}}}

\newcommand{\bi}[1]{\bibitem{#1}}
\newcommand{\fr}[2]{\frac{#1}{#2}}

\newcommand{\psl}{\slash{\!\!\!p}}
\newcommand{\cp}{\;\;\slash{\!\!\!\!\!\!\rm CP}}
\newcommand{\qq}{\langle \ov{q}q\rangle}

\newcommand{\dGd}{\bar{d}g_s(G\si) d}
\newcommand{\nc}{\newcommand}

\newcommand{\dd}{\bar{d}d}
\nc{\gone}{\bar g_{\pi NN}^{(1)}}
\nc{\gzero}{\bar g_{\pi NN}^{(0)}}
\nc{\go}{\bar g_{\pi NN}^{(1)}}
\nc{\gz}{\bar g_{\pi NN}^{(0)}}
\nc{\al}{\alpha}
\nc{\ga}{\gamma}
\nc{\de}{\delta}
\nc{\ep}{\epsilon}
\nc{\ze}{\zeta}
\nc{\et}{\eta}
\renewcommand{\th}{\theta}
\nc{\Th}{\Theta}
\nc{\ka}{\kappa}
\nc{\la}{\lambda}
\nc{\rh}{\rho}
\nc{\si}{\sigma}
\nc{\ta}{\tau}
\nc{\up}{\upsilon}
\nc{\ph}{\phi}
\nc{\ch}{\chi}
\nc{\ps}{\psi}
\nc{\om}{\omega}
\nc{\Ga}{\Gamma}
\nc{\De}{\Delta}
\nc{\La}{\Lambda}
\nc{\Si}{\Sigma}
\nc{\Up}{\Upsilon}
\nc{\Ph}{\Phi}
\nc{\Ps}{\Psi}
\nc{\Om}{\Omega}
\nc{\ptl}{\partial}
\nc{\del}{\nabla}
\nc{\ov}{\overline}
\nc{\newcaption}[1]{\centerline{\parbox{11cm}{\caption{#1}}}}
\newcommand{\qqg}{\langle \ov{q}\ga_5q\rangle}

\begin{document}

\begin{frontmatter}

\title{Electric dipole moments as probes \\ of new physics}

\author{Maxim Pospelov$^{a,b,c}$, Adam Ritz$^d$}

\address{$^a$Department of Physics, University of Guelph, Guelph,\\
Ontario N1G 2W1, Canada \\ 
$^b$Perimeter Institute for Theoretical Physics, Waterloo,
Ontario N2J 2W9, Canada \\
$^c$Department of Physics and Astronomy,
University of Victoria, Victoria,\\ British Columbia V8P 1A1, Canada\\
$^d$Theoretical Division, Department of Physics, CERN,
Geneva 23, CH-1211 Switzerland}


\begin{abstract}

We review several aspects of flavour-diagonal CP violation, focussing on the role
played by the electric dipole moments (EDMs) of leptons, nucleons, atoms and molecules,
which consitute the source of several stringent constraints on  new
CP-violating physics. We dwell specifically on the calculational aspects of 
applying the hadronic EDM constraints, reviewing in detail the application of QCD sum-rules to
the calculation of nucleon EDMs and CP-odd pion-nucleon couplings.  
We also consider the current status of EDMs in the Standard Model, and on the ensuing constraints 
on the underlying sources of CP-violation in physics beyond the Standard Model, 
focussing on weak-scale supersymmetry.

\end{abstract}

\end{frontmatter}


\section{Introduction}

The search for violations of fundamental symmetries has played a central role in the
development of particle physics in the 20$^{th}$ century. In particular, tests of the discrete symmetries,
charge conjugation $C$, parity $P$, and time-reversal $T$, have been  of paramount importance in establishing
the structure of the Standard Model (SM). Perhaps the most 
famous example was the discovery of parity violation in the
weak interactions \cite{parity}, which led to the realization that matter fields should be 
combined into asymmetric
left- and right-handed chiral multiplets, one of the cornerstones of the Standard Model. 
The observation of $CP$ violation, via the mixing of Kaons \cite{Kaon},
also subsequently provided strong evidence for the presence of three quark and lepton generations, via the
Kobayashi-Maskawa mechanism \cite{KM}, prior to direct experimental evidence for the third family.

It is interesting to recall that one of the first tests of this kind, actually pre-dating the 
discovery of parity violation in the weak interactions, was a probe of parity invariance within the -- at the 
time unknown -- theory of the strong interactions.  In 1949, Purcell and Ramsey argued, in a way that at the
time was not fully appreciated that, lacking a theory of the strong interactions, there
was no way to ``derive'' parity invariance and thus one must confirm parity 
conservation by experimental tests, or discover the lack thereof, rather than rely on a belief that 
nature ``must be symmetric''. As a probe of parity violation, Purcell and Ramsey proposed that one 
consider an intrinsic electric dipole moment of the neutron. Placed in a magnetic and electric field, 
a neutral nonrelativistic particle of spin $S$ can be described by the following Hamiltonian,
\ba
 H = - \mu {\bf B} \cdot \fr{\bf{S}}{S} - d {\bf E} \cdot \fr{\bf{S}}{S}\ .
\label{starting}
\ea
Under the reflection of spatial coordinates, $P({\bf B} \cdot \bf{S})= {\bf B} \cdot \bf{S}$, 
whereas $P({\bf E} \cdot \bf{S})= -{\bf E} \cdot \bf{S}$. The presence of a non-zero $d$
would therefore signify the existence of parity violation. It was soon realized that $d$ also 
breaks time-reversal invariance. Indeed, under time reflection, 
$T({\bf B} \cdot \bf{S})= {\bf B} \cdot \bf{S}$ and 
$T({\bf E} \cdot \bf{S})= -{\bf E} \cdot \bf{S}$. Therefore a non-zero 
$d$ may exist if and only if both parity and time reveral invariance are 
broken. Analysis of the existing experimental data on neutron scattering from spin 
zero nuclei led to the conclusion, $|d_n| < 3\times 10^{-18} e\,$cm \cite{PR}. Such a 
result probes physics at distances much shorter than the typical scale of 
nuclear froces $\sim 1$fm, or the Compton wavelength of the neutron. This initial limit on
the neutron EDM implied that $P$ and $T$ were good symmetries of the strong interactions at
percent-level precision.

It was only some 25 years later with the emergence of QCD that the possibility of 
$T$-violation (or $CP$-violation, on assuming the $CPT$ theorem) in the strong interactions had 
some theoretical underpinning. Indeed, QCD allows for the addition of a dimension-four term, known
as the $\th$-term, with a dimensionless coefficient $\th$ which, if nonzero, would signify  
the violation of both $P$ and $T$. This term is somewhat unusual, being a purely topological boundary term,
but its value determines a superselection sector in QCD \cite{theta} and its presence is 
intrinsically tied to an elegant feature of the theory, namely the mechanism via
which the mass of the $\et'$ meson is lifted well-above the scale one might naturally expect given its
apparent status as a Goldstone boson \cite{thooft}. However, were $\th\sim {\cal O}(1)$, one
would predict a neutron EDM of sufficient size to ensure that the original analysis of Purcell and
Ramsey would have detected it. In fact $\th$ is now known to be tuned to better than one part in 
$10^{9}$! This tuning is the well-known {\it strong $CP$ problem} of the
Standard Model, which has been with us for more than 25 years, and has led to interesting dynamical mechanisms
for its resolution; some of these have
important consequences and predictions for other aspects of particle physics and cosmology.

The required tuning of this $CP$-odd parameter in QCD comes into sharp focus when we put QCD 
into its rightful place 
within the Standard Model, which necessarily means coupling it to the electroweak sector and massive 
quarks in particular. In this case the physical value of $\th$ acquires a contribution from the overall phase
of the quark mass matrix. In this sense the strong $CP$ problem can be phrased as the absence, to high
precision, of {\it flavour-diagonal} $CP$-violation within the Standard Model. This situation could not 
contrast more
strongly with the situation in the flavour-changing sector, which is where all currently 
observed $CP$-violating effects reside. Indeed, the original discovery of $CP$-violation
in the system of neutral Kaons \cite{Kaon}, can be explained within this sector through the elegant
and indeed rather minimal model of Kobayashi and Maskawa, which links $CP$-violation to the single 
physical phase in the unitary CKM mixing matrix describing transitions between
the three generations of quarks \cite{KM}. This picture has
recently received significant support -- indeed essential confirmation -- through experiments using
neutral $B$ mesons \cite{sin2beta}. In contrast to $\th$, the phase in the CKM mixing matrix requires
no tuning at all -- its effects are nicely masked in the appropriate channels by the flavour structure of
the Standard Model. Indeed, it turns out that the predictions for any $CP$-violating effect in the 
flavour-conserving 
channel induced by CKM mixing are minuscule, thus denying any hopes of detecting the experimental manifestation 
of CKM physics in these channels in the foreseeable future.

Searches for flavour-diagonal $CP$-violation, while insensitive to the CKM phase, thus inherit on the flip-side
the status as one of the unique, essentially ``background'' free, probes of new physics. Electric dipole moments,
through continuous experimental development since the work of Purcell and Ramsey, remain our most 
sensitive probes of this sector. All existing searches have failed to detect any intrinsic EDM, and indeed the 
precision to which EDMs are now known to vanish is remarkable, and sufficient to render them some of the 
most important precision tests of the Standard Model. In this more general context,
the strong $CP$ problem, associated with the tuning of $\th$, becomes just the most highly tuned example among
many possible $CP$-odd operators which could contribute to the observable EDMs of nucleons, leptons, 
atoms and molecules. Anticipating the presence of such $CP$-odd sources is not without motivation. 
Indeed, one of the strongest motivations comes from cosmology, where the success of the inflationary scenaro, 
together with the observed cosmological dominance of baryons over antibaryons, suggests that 
a non-zero baryon number was generated dynamically in the early Universe. According to the 
Sakharov criteria \cite{Sakharov}, this requires a source of $CP$-violation, and within the Standard
Model, neither the Kobayashi-Maskawa phase nor the $\th$--term can create conditions that would lead 
to the generation of an appreciable net baryon number. This strongly suggests the presence of another, 
yet to be discovered, source of $CP$ violation in nature. Although exceptions exist,
e.g. the leptogenesis scenario, EDMs generally provide a highly sensitive diagnostic for these new 
$CP$-odd sources. 

The second prominent motivation arises from theoretical prejudices about the physics of
the Fermi scale, i.e. the mechanism for electroweak symmetry breaking, currently the focus of
intense theoretical and experimental work. There are several theoretical motivations to believe 
that new physics, beyond the SM Higgs boson, should become apparent at, or just above, this scale, 
with weak-scale supersymmetry (SUSY) being a prominent example. Flavour-diagonal $CP$-violation 
constitutes a powerful probe of these scales, since any new physics need not provide the same 
flavour-dependent suppression factors as does the SM, while the SM itself
constitutes a negligible background. These precision tests are thus highly complementary to direct searches
at colliders. A rough estimate, based on the decoupling of new physics as the inverse square of its 
characteristic 
energy scale $\Lambda$, currently gives us the possibility to probe an order one $CP$-violating source at 
up to $\Lambda\sim 10^{6}$ GeV. In many weakly coupled theories, such as SUSY, this scale is somewhat 
lower, but often is still beyond the reach of existing and/or projected colliders. 
As with the link between the Kobayashi-Maskawa mechanism and the three-generation structure, one might hope
that flavour-diagonal $CP$-violation, or perhaps the lack thereof, will tell us something profound about 
the matter sector.

The level of experimental precision achieved in EDM searches has improved dramatically since the
early work of Purcell and Ramsey, and has been broadened to many atomic and nuclear quantities.
Indeed, following significant progress throughout the past decade, the EDMs of the neutron \cite{n}, 
and of several 
heavy atoms and molecules \cite{Tl,Hg,TlF,Xe,Cs,YbF} have been measured to vanish to 
remarkably high precision. From the present standpoint, it is convenient to classify the EDM searches 
into three main 
categories, distinguished by the dominant physics which would induce the EDM, at least 
within a generic class of models. These categories are: the EDMs of paramagnetic atoms and 
molecules; the EDMs of diamagnetic atoms; and the EDMs of hadrons, and nucleons in particular. 
For these three categories, the experiments that currently champion the best bounds on 
$CP$-violating parameters are the atomic EDMs of thallium and mercury and that of the neutron,
as listed in Table~1.

\begin{table}[t]
\begin{center}
\newcaption{Current constraints within three representatve classes of EDMs}
\begin{tabular}{||c|c|c||}
\hline
  Class & EDM & Current Bound  \\
  \hline
  Paramagnetic & $^{205}Tl$ & $|d_{\rm Tl}| < 9 \times 10^{-25} e\, {\rm cm}$ (90\% C.L.) \cite{Tl}  \\
  Diamagnetic & $^{199}Hg$ & $|d_{\rm Hg}| < 2   \times 10^{-28}  e\, {\rm cm}$ (95\% C.L.) \cite{Hg}  \\
  Nucleon & $n$ & $|d_n|  <  6\times 10^{-26} e\, {\rm cm}$ (90\% C.L.)  \cite{n} \\
  \hline
\end{tabular}
\end{center}
\label{explimit}
\end{table}

The upper limits on EDMs obtained in these experiments can be translated into tight 
constraints on the $CP$-violating physics at and above the electroweak scale, with each category of
EDM primarily sensitive to different $CP$-odd sources. For example, the neutron EDM can be 
induced by $CP$ violation in the quark sector, while paramagnetic EDMs generally result from 
$CP$ violating sources that induce the electron EDM. Despite the apparent difference in the 
actual numbers in Table~\ref{explimit}, all three limits on $d_n$, $d_{\rm Tl}$, and $d_{\rm Hg}$ 
actually have comparable sensitivity to fundamental $CP$ violation, 
e.g. superpartner masses and $CP$-violating phases, and thus play complementary roles in constraining
fundamental $CP$-odd soources. This fact can be explained by the way the so-called 
Schiff screening theorem \cite{Schiff} is violated in paramagnetic and diamagnetic atoms.
The Schiff theorem essentially amounts to the statement that, in the nonrelativistic limit
and treating the nucleus as pointlike, the atomic EDMs will vanish due to screening of the
applied electric field within a neutral atom. The paramagnetic and diamagnetic EDMs result
from violations of this theorem due respectively to relativistic and finite-size effects,
and in heavy atoms such violation is maximized. 
For heavy paramagnetic atoms, {\i.e.} atoms with 
non-zero electron angular momentum, relativistic effects 
actually result in a net enhancement of the atomic EDM over the electron EDM. 
For diamagnetic species, the Schiff screening 
is violated due to the finite size of the nucleus, but this is a weaker effect
and the induced EDM of the atom is suppressed relative to the EDM of the nucleus itself. These factors
equilibrate the sensitivities of the various experimental constraints in Table~\ref{explimit}
to more fundamental sources of $CP$ violation.

In this paper we will review in detail the calculational aspects of applying the current bounds on EDMs to constrain
new $CP$-violating sources. In order to make the discussion as systematic as possible, we 
will proceed by working our way upwards in energy scale, using several effective 
$CP$-odd Lagrangians at the relevant thresholds. 
In the next section, we begin by discussing the current status of the experimental constraints within 
three generic classes, namely the neutron EDM, and the EDMs of paramagnetic and diamagnetic atoms, and 
describing the
contributions to these EDMs at the nuclear scale. We then move to the QCD scale and introduce an effective
$CP$-odd effective quark-gluon Lagrangian which plays an important role in the subsequent analysis. The
leading term in this effective theory, of dimension four, is the $\th$-term and we breifly review the 
strong $CP$-problem and some of its proposed resolutions. We then turn in Section~3 to QCD computations of
the EDMs, and dwell on some of the calculational aspects which are currently some of the 
major sources of uncertainty in the application of EDM constraints (for a detailed discussion of 
many other aspects we refer the reader to \cite{KL}). In Section~4, we turn to the 
generation of these observables within specific models of $CP$-violation,
reviewing first the significant sources of suppression within the Standard Model, and then focussing on weak 
scale supersymmetry, and the MSSM in particular, as the source of new physics at the electroweak scale. 
We discuss the generic constraints that EDMs impose on 
combinations of $CP$-violating parameters in the SUSY-breaking sector, and also explore some additional effects
which may arise in special parameter regimes. We also emphasize the stringent EDM constraints on 
combined sources of $CP$- and flavour-violation in more general models. Finally, we conclude in Section~5 
with an outlook on future experimental
and theoretical developments.

\section{EDMs as probes of $CP$ violation}

The majority of EDM experiments are performed with matter as opposed to
anti-matter. Therefore, the conclusion about the relation between $d$ and $CP$ violation relies 
on the validity of the $CPT$ theorem. The interaction $d \bf{E}\cdot \bf{S}$ for a 
spin 1/2 particle then has the following relativistic generalization
\ba
\label{relativ}
H_{\rm T,P-odd} =  - d {\bf E} \cdot \fr{\bf{S}}{S} \qquad \longrightarrow \qquad
{\cal L} = -d \fr{i}{2}\overline{\psi} \sigma^{\mu\nu}\gamma_5 \psi 
F_{\mu\nu}.
\ea
Parenthetically, it is worth remarking that the precision of EDM experiments has now reached 
a level sufficient to provide competetive tests of $CPT$ invariance, 
since one
can also consider a $CP$-even, but $CPT$-odd, relativisitic form of $d \bf{E}\cdot \bf{S}$,
namely ${\cal L} = d \overline{\psi} \gamma^\mu\gamma_5 \psi F_{\mu\nu}n^\nu$, with
a preferred frame $n^\nu = (1,0,0,0)$, which spontaneously breaks Lorentz invariance and $CPT$.

The problem of calculating an observable EDM from the underlying 
$CP$ violation in a given particle physics model can be conveniently 
separated into different stages, depending on the characteristic energy/momentum scales. 
At each step the result can be expressed as an effective Lagrangian in terms of the
light degrees of freedom with Wilson coefficients that encode information about 
$CP$ violation at higher energy scales. As usual in effective field theory, it is convenient to 
classify all possible effective $CP$ violating operators in terms of their 
dimension, with the operators of lowest dimension usually leading to the largest contributions. 
This logic may need to be refined if symmetry requirements imply that certain operators are 
effectively of higher dimension than naive counting would suggest. This is actually the case for
certain EDM operators due to gauge invariance, as discussed in more detail below.

We will present this analysis systematically in order of increasing energy scale, working our way upwards 
in the dependency tree outlined in Fig.~1, which allows us to
remain entirely model-independent until the final step where some high-scale model of $CP$ violation
can be imposed and then subjected to EDM constraints.

\begin{figure}[t]
 \centerline{%
   \includegraphics[width=10cm]{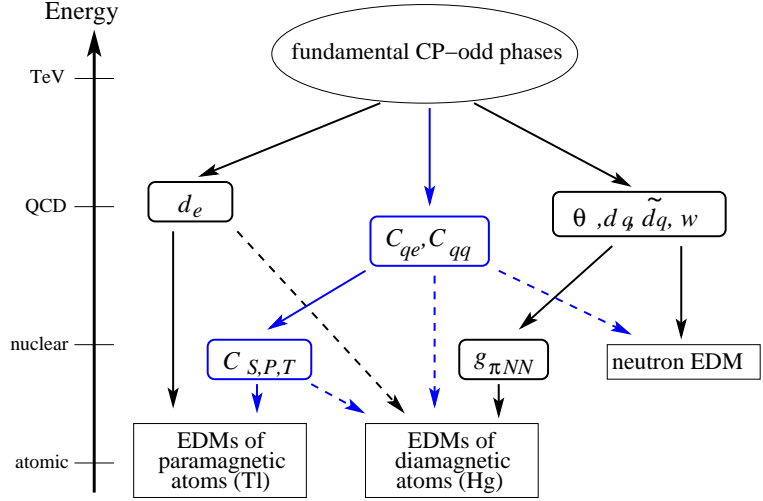}%
         }
\vspace{0.1in}
 \caption{\label{schemetb}A schematic plot of the hierarchy of scales between the CP-odd
 sources and three generic classes of observable EDMs. The dashed lines indicate generically
 weaker dependencies.}
\end{figure}

\subsection{Observable EDMs}
  
Let us begin by reviewing the lowest level in this construction, namely the precise relations between
observable EDMs and the relevant $CP$-odd operators at the nuclear scale. At leading order, such effects may be
quantified in terms of EDMs of the constituent nucleons, $d_n$ and $d_p$ (where the 
neutron EDM is already an observable), the EDM of the electron $d_e$, and $CP$-odd electron-nucleon
and nucleon-nucleon interactions. In the relevant channels these latter interactions are 
dominated by pion exchange, and thus we must also 
consider the $CP$-odd pion-nucleon couplings $\bar{g}_{\pi NN}$ which can be induced by  
$CP$-odd interactions between quarks and gluons. 
To be more explicit, we write down the relevant 
$CP$-odd terms at the nuclear scale,
\ba
 {\cal L}_{eff}^{\rm nuclear} = {\cal L}_{\rm edm} + {\cal L}_{\pi NN}+ {\cal L}_{eN},
  \label{Lnuc}
\ea
which can be split into terms for the nucleon (and electron) EDMs,
\ba
 {\cal L}_{\rm edm} =  - \frac{i}{2} \sum_{i=e,p,n} d_i\ 
  \overline{\psi}_i (F\sigma)\gamma_5 \psi,
\ea
the $CP$-odd pion nucleon intercations,
\ba
 {\cal L}_{\pi NN } &=& \bar g_{\pi NN}^{(0)}\bar N\tau^aN\pi^a
+\bar g_{\pi NN}^{(1)}\bar NN\pi^0 \nonumber\\
 && \;\;\;\;\;\;\;\;\;\;\;+\bar g_{\pi NN}^{(2)}(\bar N\tau^aN\pi^a - 3 \bar N\tau^3N\pi^0),
 \label{LpiNN}
\ea
and finally $CP$-odd electron-nucleon couplings,
\ba
 &&{\cal L}_{eN} = C_S^{(0)} \bar e i \gamma_5 e \bar NN +  C_P^{(0)} \bar e e \bar N i\gamma_5 N
+ C_T^{(0)} \epsilon_{\mu\nu\alpha\beta}
\bar e\sigma^{\mu\nu} e \bar N \sigma^{\alpha\beta} N \;\;\;\;\;\;\;\;\;\;\;\;\;\;\nonumber\\
 &&\;\;\;\;+C_S^{(1)} \bar e i \gamma_5 e \bar N\ta^3N +  C_P^{(1)} \bar e e \bar N i\gamma_5\tau^3 N
+ C_T^{(1)} \epsilon_{\mu\nu\alpha\beta}
\bar e\sigma^{\mu\nu} e \bar N \sigma^{\alpha\beta} \tau^3N.\;\;\;\;\;\;\;
\label{eN}
\ea
In certain rare cases, $CP$-odd nucleon-nucleon forces are not mediated by pions, in 
which case the effective Lagrangian must be extended by a variety of {\em contact} terms  
e.g. $\bar N N \bar N i\gamma_5 N$, and the like. 

The dependence of the observable EDMs on the corresponding Wilson coefficients relies on atomic and
nuclear many-body calculations which would go beyond the scope of this review to cover here (see the reviews
\cite{KL,fgreview} for further details). However, 
we will briefly summarize the current status of these calculations, before turning to our major focus which
is the calculation of these coefficients in terms of higher scale $CP$-odd sources.

As alluded to earlier on, it is convenient to split the discussion into three parts, corresponding
roughly to the three classes of observable EDMs which currently provide constraints at a similar level of
precision; namely: EDMs of paramagnetic atoms and molecules, EDMs of diamagnetic atoms, and the neutron 
EDM.

\bigskip
$\bullet$ {\it EDMs of paramagnetic atoms -- thallium EDM}
\bigskip

Paramagnetic systems, namely those with one unpaired electron, are primarily sensitive to the 
EDM of this electron. At the nonrelativistic level, this is far from obvious due to the
Schiff shielding theorem which implies, since the atom is neutral, that any applied electric field
will be shielded and so an EDM of the unpaired electron will not induce an atomic EDM. Fortunately, this
theorem is violated by relativistic effects. In fact, it is violated strongly for atoms with a large atomic 
number, and even more strongly in molecules which can be polarised by the applied field. For atoms, 
the parameteric enhancement of the electron EDM is given by \cite{sandars,LiuKelly,fgreview},
\ba
 d_{\rm para}(d_e) \sim 10 \frac{Z^3\al^2}{J(J+1/2)(J+1)^2}d_e,
\ea
up to numerical ${\cal O}(1)$ factors, with $J$ the angular momentum and $Z$ the atomic number.
This enhancement is significant, and for large $Z$, the applied field can be enhanced be a 
factor of a few hundred within the atom. This feature explains why atomic systems provide such a powerful probe of
the electron EDM, since the ``effective'' electric field can be much larger than one could actually
produce in the lab.

Although the electron EDM is the predominant contributor to any paramagnetic EDM is most models,
one should bear in mind that other contributions may also be significant in certain regimes. In particular,
significant $CP$-odd electron-nucleon couplings may also be generated, 
due for example to $CP$ violation in the Higgs sector. 
Among these couplings, $C_S$ plays by far the most important 
role for paramagnetic EDMs 
because it couples to the spin of the electron and is enhanced by the
large nucleon number in heavy atoms.

Among various paramagnetic systems, the EDM of the thallium atom currently provides
the best constraints on fundamental $CP$ violation.  A number of atomic calculations \cite{LiuKelly,MP1,MP2} 
(see also Ref. \cite{KL} for a more complete list) have 
established the relation between the EDM of 
thallium, $d_e$, and the CP-odd electron-nucleon interactions $C_S$:
\ba 
d_{\rm Tl} = -585 d_e -   e\ 43 ~{\rm GeV}
 \times ( C_S^{(0)} - 0.2 C_S^{(1)}).
\label{dtl}
\ea
with $C_S$ expressed in isospin components. 
The relevant atomic matrix elements are known to within 
$10-20\%$ \cite{fgreview}.

As we discuss later on, current experimental work is focussing on the use of paramagnetic 
molecules, e.g. YbF and PbO \cite{YbF,PbO}, which can provide an 
even larger enhancement of the applied field due to polarization 
effects, have better systematics,
and may bring significant progress in measuring/constraining $d_e$ and $C_S$. 

\bigskip
$\bullet$ {\it EDMs of diamagnetic atoms -- mercury EDM}
\bigskip

EDMs of diamagnetic atoms, i.e. atoms with total electron angular momentum equal to zero, 
also provide an important test of $CP$ violation \cite{KL}. In such systems the Schiff shielding argument again 
holds to leading order. However, in this case it is violated not by relativistic effects but by finite
size effects, namely a net misalignment between the distribution of charge and EDM (i.e.
first and second moments) in the nucleus of a large atom (see e.g. \cite{fgreview} for a review).
However, in contrast to the paramagnetic case, this is a rather subtle effect and the induced atomic
EDM is considerably suppressed relative to the underlying EDM of the nucleus.

To leading order in an expansion around the point-like approximation for the nucleus, the contributions arise from
an octopole moment (which is only relevant for states with large spin, and will not be relevant for the
cases considered here), and the Schiff moment $\vec{S}$, which contributes to the electrostatic potential,
\ba
 V_{E} = 4\pi \vec{S}\cdot \vec{\nabla}\de(\vec{r}). 
\ea
$CP$-odd nuclear moments, such as $\vec{S}$, can 
arise from intrinsic EDMs of the constituent nucleons and also $CP$-odd
nucleon interactions. It turns out that the latter source tends to dominate in diamagnetic atoms
and thus, since such interactions are predominantly due to pion exchange, we can ascribe the 
leading contribution to $CP$-odd pion nucleon couplings $\bar{g}_{\pi NN}^{(i)}$ for $i=0,1,2$ corresponding
to the isospin.

There are of course various additional contributions, which are generically subleading, but may become
important in certain models. Schematically, we can represent the EDM in the form
\ba
 d_{\rm dia} = d_{\rm dia}(S[\bar{g}_{\pi NN},d_N],C_S,C_P,C_T,d_e),
\ea
where we note that electron-nucleon interactions may also be significant, as is the electron EDM itself
\cite{KL} (although in practice the electron EDM tends to be more strongly constrained by limits
from paramagentic systems and thus is often neglected).

Currently, the strongest constraint in the diamagnetic sector comes from the bound on the EDM of
mercury -- at the atomic level, this is in fact the most precise EDM bound in existence. As should be apparent
from the above discussion, computing the dependence of $d_{\rm Hg}$ on the underlying $CP$-odd sources is 
a nontrivial problem requiring input from QCD and nuclear and atomic physics. In particular,
the computation of $S(\bar{g}_{\pi NN})$ is a nontrivial nuclear many-body problem, and has recently been
reanalyzed with the result \cite{DS},
\ba
 S(^{199}{\rm Hg}) = -0.0004 g\bar{g}^{(0)} -0.055 g\bar{g}^{(1)}+ 0.009 g\bar{g}^{(2)} \; e\; {\rm fm}^3,
\ea
where $g=g_{\pi NN}$ is the $CP$-even pion-nucleon coupling, and $\bar{g}^{(i)}=\bar{g}_{\pi NN}^{(i)}$
denote the $CP$-odd couplings. The isoscalar and isotensor couplings have been significantly reduced
relative to previous estimates, while the isovector coupling -- which generically turns out to be most
important -- has been less affected (within a factor 2). 
This nonethless provides some indication of the difficulties inherent
in the calculation, and makes precision estimates more difficult. Moreover, it is worth
noting that the suppression of the 
overall coefficient in front of $g\bar{g}^{(0)}$ below $O(0.01)$  
is the result of mutual cancellation between several contributions of comparable size, 
and therefore is in some sense accidental and may not hold in future refinements of these
nuclear calculations. 

Putting the pieces together, we can write the mercury EDM in the form,
\ba
 d_{{\rm Hg}} &=& -(1.8 \times 10^{-4}\, {\rm GeV}^{-1}) e\,\bar{g}_{\pi NN}^{(1)} + 10^{-2} d_e\nonumber\\
&&\;\;\;\;\;\;\;+ (3.5\times 10^{-3}\, {\rm GeV}) e\, C_S^{(0)}, 
\ea
where we have limited attention to the isovector pion-nucleon coupling and  $C_S$ 
which turns out to be the most important for $CP$ violation in supersymmetric models.

\bigskip
$\bullet$ {\it Neutron EDM}
\bigskip

The final class to consider is that of the neutron itself, whose EDM can be searched for directly
with ultracold neutron technology, and 
currently provides one of the strongest constraints on new $CP$-violating physics.
In this case, there is clearly no additonal atomic or nuclear physics to deal with, and we must turn
directly to the next level in energy scale, namely the use of QCD to compute the dependence of $d_n$
on $CP$-odd sources at the quark-gluon level. This statement also applies to many of the other quantities
we have introduced thus far, including in particular the $CP$-odd pion-nucleon coupling. Indeed,
it is only paramagnetic systems that are partially immune to QCD effects, although even there
we have noted the possible relevance of electron-nucleon interactions.

\subsection{The structure of the low energy Lagrangian at 1~GeV} 

The effective CP-odd flavour-diagonal Lagrangian normalized at 1 GeV, which is taken to be the
lowest perturbative quark/gluon scale, plays a special role in EDM calculations. 
At this scale, all particles other than 
the $u,d$ and $s$ quark fields, gluons, photons, muons and electrons can be considered 
heavy, and thus integrated out. As a result, one can construct an effective 
Lagrangian by listing all possible $CP$-odd operators in order of increasing dimension,
\ba
{\cal L}_{\rm eff} = {\cal L}_{\rm dim=4} + {\cal L}_{\rm dim=5} +{\cal L}_{\rm dim=6} +\cdots.
\label{456}
\ea
There is only one operator at dimension 4, the QCD theta term,
\ba
\label{thetaterm}
{\cal L}_{\rm dim=4} =\frac{g_s^2}{32\pi^{2}}\ \bar\theta
G^{a}_{\mu\nu} \widetilde{G}^{\mu\nu , a} ,
\ea
where on account of the axial U(1) anomaly, the physical value of $\th$ -- denoted $\bar\th$ -- also
includes the overall phase of the quark mass matrix, 
\ba
 \bar\th = \th + {\rm Arg\, Det}M_q.
\ea
The anomaly can be used to shuffle contributions between the $\th$-term and imaginary quark
masses, but only the combination $\bar\th$ is physical and we choose to place it in front of
$G\tilde{G}$ taking Det$M_q$ to be real. It should be apparent that if any of the quarks were massless,
we could then rotate $\th$ away and it would have no physical consequences.

At the dimension five level, there are (naively) several operators: EDMs of light 
quarks and leptons and color electric dipole moments of the light quarks,
\ba
\label{manyedms}
{\cal L}_{\rm dim=5} = - \frac{i}{2} \sum_{i=u,d,s,e,\mu} d_i\ 
\overline{\psi}_i (F\sigma)\gamma_5 \psi_i  -
\frac{i}{2} \sum_{i=u,d,s} 
\widetilde{d}_i\ \overline{\psi}_i g_s (G\sigma)\gamma_5\psi_i,\;\;\;\;\;\;\;\;
\ea
where $(F\si)$ and $(G\si)$ are a shorthand notation for 
$F_{\mu\nu} \sigma^{\mu\nu}$ and $G^a_{\mu\nu}t^a \sigma^{\mu\nu}$. 

In fact, in most models these operators are really dimension-six operators in disguise. The reason is that,
if we proceed in energy above the electroweak scale and assume the system restores SU(2)$\times U(1)$ 
as in the Standard Model, gauge invariance ensures that these operators must include a Higgs field 
insertion \cite{signets}. Indeed, were we to write the basis of down quark EDMs and CEDMs above the 
electroweak scale, we should specify the following list of dimension six operators \cite{signets},
\ba
 {\cal L}_{\rm ``dim=5''}^{\rm EW} &=& \frac{i}{2\sqrt{2}} \bar{Q}_L \left[ 
 2 d^{\rm EW}_1 (B\si) + d^{\rm EW}_2 \ta^i (W^i\si) \right.\nonumber\\
   && \;\;\;\;\;\;\;\;\;\;\;\;\;\;\;\;\;+ \left. d^{\rm EW}_2 \la^a (G^a\si)\right](\Ph/v) D_R
  +{\rm h.c.}, \label{ginvt}
\ea
which are defined in terms of left-handed doublets $Q_L=(U,D)_L$ and right-handed singlets $D_R$
and the Higgs doublet $\Ph$, and in terms of the U(1), SU(2), and SU(3) field strengths 
$B_{\mu\nu}$, $W_{\mu\nu}^i$ and $G_{\mu\nu}^a$. 

The lesson we draw from (\ref{ginvt}) with regard to EDMs is that, if generated, these operators 
must be proportional to the Higgs v.e.v. below the electroweak scale, and consequently 
must scale at least as $1/M^2$ for $M\gg M_W$.
In practice, this feature can also be understood in most models by going to a chiral basis, where we see
that these operators connect left- and right-handed fermions, and thus require a chirality flip.
This is usually supplied by an insertion of the fermion mass, i.e. $d_f \sim m_f/M^2$, again implying
that the operators are effectively of dimension six. 

Consequently, for consistency we should also proceed at least to dimension six where
we encounter the $CP$-odd three-gluon Weinberg operator and a host of possible four-fermion interactions,
$(\bar \psi_i \Ga\psi_i)(\bar \psi_j i \Ga\gamma_5 \psi_j)$, where $\Ga$ denotes 
several possible scalar or tensor Lorentz structures and/or gauge structures, which are contracted
between the two bilinears. We limit our attention to a small subset of 
the latter that will be relevant later on, 
\ba
{\cal L}_{\rm dim=6} = \frac{1}{3} w\  f^{a b c} G^{a}_{\mu\nu} \widetilde{G}^{\nu \beta , b}
G^{~~ \mu , c}_{\beta} + \sum_{i,j} C_{ij} ~ (\bar \psi_i \psi_i)(\bar \psi_j i \gamma_5 \psi_j) +\cdots
  \;\;\;\;\;\;\;\;\;
\label{manydim6}
\ea
In this formula, the operators with $C_{ij}$ are summed over all light fermions.
Going once again to a chiral basis, we can argue as above that the four-fermion
operators, which require two chirality flips, are in most models effectively of dimension eight.
Nonetheless, in certain cases they may be non-negligible.

\subsection{The strong $CP$ problem}

The leading dimension-four term in the $CP$-odd Lagrangian given in Eq.~(\ref{thetaterm})
has a special status,
in that it is a marginal operator, unsuppressed by any heavy scale. It is also a total derivative -- 
we can write $G\tilde{G} = \ptl_\mu {\cal K}^{\mu}$ with ${\cal K}^{\mu}$ the Chern-Simons current -- and
thus plays no role in perturbation theory. However, ${\cal K}^{\mu}$ is not invariant under so-called
{\it large} gauge transformations and thus one may expect that the $\th$-term becomes relevant at 
the nonperturbative level. That it does so can be argued at the semi-classical level using instanton methods,
and more generally can be understood within QCD via this relation to the U(1) problem. In particular, 
we note that the same operator $G\tilde{G}$ arises as the $\th$-term in the Lagrangian, and 
also as an anomaly for the axial U(1) current ${\cal J}_A^{\mu}$, i.e. for massless quarks,
\ba
 \ptl_\mu{\cal J}^\mu_A = \frac{\al_s}{2\pi} G_{\mu\nu}^a\tilde{G}^{\mu\nu a}.
\ea
This leads to an intrinsic link between two physical phenomena: namely the $\th$-dependence of physical
quantities, and the absence of a light pseudo-Goldstone boson associated with spontaneous breaking
of the axial current ${\cal J}_A^{\mu}$ \cite{thooft} (the corresponding state, the $\et'$ is instead rather heavy, 
$m_{\et'}\gg m_{\pi}$). Although it would take us too far afield to review the story of this link in detail
(see e.g. \cite{thooft,crewther,witten,veneziano,svzcp,de}), let us note that in the 
large $N$ limit, as discussed 
by Witten and Veneziano \cite{witten,veneziano}, use of the anomaly equation leads to a 
simple relation that exemplifies this connection,
\ba
 m_{\et'}^2 = \frac{4 N_f}{f_\pi^2} \left( \frac{d^2 E}{d\th^2}\right)^{\rm YM}_{\th=0}, \label{met'}
\ea
where $N_f$ is the number of flavours. This relation expresses the $\et'$ mass in terms of the $\th$-dependence of the 
vacuum energy in a theory with no light quarks. 

In turn, if we now take for granted that $m_{\et'}\gg m_{\pi}$, use of the anomaly relation
allows precise calculations of the $\th$-dependence of physical observables \cite{crewther,svzcp}. In particular,
one can obtain an expression for the $\th$-dependence of the vacuum energy $E(\th)$. At leading order in $\bar\th$,
\ba
 E(\bar\th) = \frac{1}{2}\bar\th^2 \ch(0) = -\frac{i}{8\pi^2}\bar\th^2 \lim_{k\rightarrow 0}
   \int d^4 x e^{ipx}\left\langle \frac{\al_s}{2\pi} G\tilde{G}(x), \frac{\al_s}{2\pi} G\tilde{G}(0) 
   \right\rangle,
\ea 
where $\ch(0)$ is known as the topological susceptibility. Making use of the anomaly relation and 
assuming $m_{\et'}\gg m_{\pi}$, this may be evaluated as \cite{svzcp}
\ba
 E(\bar\th) = -\frac{1}{2}\bar\th^2 m_* \qq +{\cal O}(\bar\th^4,m_*^2),\label{Energy}
\ea
where $\qq$ is the quark vacuum condensate and $m_*$ is the reduced quark mass, given by 
\ba
   m_* = \frac{m_um_d}{m_u+m_d}, 
\ea
in two-flavour QCD. This dependence on the reduced quark mass can be straightforwardly understood on
recalling that $\bar\th$ becomes unphysical as soon as any quark eigenstate becomes massless. Indeed, we 
see that the result is essentially fixed up to an order-one coefficient by the dictates of the 
anomalous Ward identity. On general grounds, we would expect $E(\bar\th)\sim \bar\th^2 m_* \La_{\rm had}^3$,
where $\La_{\rm had}$ is the characteristic hadronic scale required on dimensional grounds, which the
calculation above identifies with the quark condensate. 

The quadratic dependence of the vacuum energy on $\bar\th$, since its determined effectively by a two-point
function, implies that generic $CP$-odd observables will inherit a leading linear dependence on $\bar\th$.
In particular, although we will discuss a more detailed calculation in the next section, 
we can obtain a similar order of magnitude estimate for the neutron EDM,
\ba
 d_n \sim e\, \frac{\bar\th m_*}{\La_{\rm had}^2} \sim \bar\th \cdot (6 \times 10^{-17})\ e\, {\rm cm},
\ea
where we identified $\La_{\rm had}=m_n$ and used conventional values for the light quark masses. The experimental
bound then translates into the limit,
\ba
 |\bar\th| < 10^{-9}.
\ea
This remarkable degree of tuning in the value of $\bar\th$ then constitutes 
the {\it strong $CP$ problem}. It is aggravated 
by the fact that $\bar\theta$ is a dimensionless parameter, and thus can receive
corrections from unspecified sources of CP violation at an arbitrarily high scale.

If we discard the possibility that this tuning is simply accidental, and search for a theoretical
explanation for why $\bar\th$ is very small, then we find that the existing
theoretical attempts to solve the strong $CP$ problem can be divided into those that are based either on 
continuous symmetries or on spontaneously broken discrete symmetries. To some extent, these two possibilities
can also be motivated by two extreme reference points, namely when $\bar{\th}$ is either fully rotated to sit 
in front of $G\tilde{G}$, or to manifest itself
as an overall phase of the quark mass matrix. Although inherently basis-dependent, the former viewpoint
suggests that $\th$ is essentially tied to the gluonic structure of QCD, while the latter
emphasizes instead its links to the flavour sector.

\subsubsection{Dynamical relaxation of $\bar\theta$}

The energy of the QCD vacuum as a function of $\bar\theta$ (\ref{Energy}) has a minimum 
at $\bar\theta=0$. Thus the relaxation of the $\th$-parameter to zero is possible if one promotes 
it to a dynamical field, called the {\em axion} \cite{PQ,WW,w}. This is motivated by the
assumption that the Standard Model, augmented by appropriate additional fields, 
admits a chiral symmetry U(1)$_{\rm PQ}$, acting on states charged under SU(3)$_c$.
When this symmetry is spontaneously broken at a necessarily high scale $f_a$, a pseudoscalar 
Goldstone boson -- the axion -- survives as the only low energy manifestation. Symmetry dictates
that the essential  components of the axion Lagrangian are very simple, 
\ba
{\cal L}_a = \fr 12 \partial_\mu a \partial^\mu a + \fr{a(x)}{f_a} \fr{\alpha_s}{8\pi} G\tilde G,
\label{axion1}
\ea
leading to a field-dependent shift of $\th$,
\ba
 \bar\th \rightarrow \bar\th + \frac{a}{f_a}. \label{thshift}
\ea
If the effects of non-perturbative QCD are ignored, 
this Lagrangian possesses a symmetry, $a\rightarrow a+$const, and $a$ is a massless field
with derivative couplings to the SM fields, 
{\em i.e.} $ \partial_\mu a \bar \psi \gamma_\mu\gamma_5 \psi$, that are not important 
for the solution of the strong $CP$ problem.
 
Below the QCD scale, one finds that U(1)$_{\rm PQ}$ is explicitly broken by the 
chiral anomaly, and thus the axion is in reality a {\it pseudo}-Goldstone boson and acquires a
potential. The form of this potential can be read directly from our earlier discussion of 
the $\th$-dependence of the vacuum energy, namely $E(\th) \sim \ch(0)\th^2/2 +\cdots$.
Accounting for the shift in (\ref{thshift}), the effective axion Lagrangian becomes,
\ba
{\cal L}_a^{\rm eff} = \fr 12 \partial_\mu a \partial^\mu a - \frac{1}{2}\chi(0)\left(\bar\theta 
   + \frac{a}{f_a}\right)^2.
   +\cdots, \label{axion2}
\ea
We see from (\ref{thshift}) that the vacuum expectation value of the axion 
field $\langle a \rangle  $ renormalizes the value of $\bar\theta$  so that all observables depend 
on the $(\bar \theta +\langle a\rangle /f_a)$ combination. At the same time, such a combination must vanish
in the vacuum as it minimizes the value of the axion potential in (\ref{axion2}). This dynamical relaxation
then solves the strong $CP$ problem. This cancellation mechanism works independently of the 
``initial'' value of $\bar\theta$, which is why it is very appealing. However, the excitations around 
$\langle a\rangle$ correspond to a massive axion particle with 
\ba
 m_a \sim \frac{1}{f_a}|\chi(0)|^{1/2},
\ea
a formula analogous to that discussed earlier for $\et'$ (\ref{met'}). For large $f_a$ the axion is
very light and thus has significant phenomenological consequences. Indeed the negative results of direct and indirect 
searches for ``invisible axions'' \cite{ksvz,zdfs}, where $f_a$ is a free scale as we have been discussing here, 
has now imposed a rather large bound, $f_a > 10^{10}$ GeV, while cosmological constraints imply, on the contrary, 
that it cannot be too much larger than this (see {\it e.g.} \cite{pdg}).

An aspect of the axion mechanism that is perhaps not stressed as often as it should be is that
there can be other contributions to the axion potential which shift its minimum away from 
$(\bar{\th}+a/f_a)=0$ \cite{BU}.
In particular, in the analysis above, we included only the leading term corresponding to the vacuum energy.
However, if there are other $CP$-odd operators ${\cal O}_{\rm CP}$ present at low scales, QCD effects 
may also generate terms linear in $\th$ via nonzero mixed correlators of the form
\ba
 \ch_{{\cal O}_{\rm CP}}(0) = -i  \lim_{k\rightarrow 0} \int d^4 x e^{ik\cdot x} 
\langle 0| T( G\tilde{G}(x), {\cal O}_{\rm CP}(0))| 0\rangle.
\ea
An example of this type is the quark chromoelectric dipole moment, 
${\cal O}_{\rm CP} = \tilde{d}_q \bar{q} G\si\ga_5 q$, appearing in (\ref{manyedms}).
The axion potential is then modified, 
\ba
 {\cal L}_a^{\rm eff} = \fr 12 \partial_\mu a \partial^\mu a - \ch_{{\cal O}_{\rm CP}}(0)\left(\bar\theta 
   + \frac{a}{f_a}\right)- \frac{1}{2}\chi(0)\left(\bar\theta + \frac{a}{f_a}\right)^2
   +\cdots, \label{axion3}
\ea
and exhibits a minimum shifted from zero. The size of this {\it induced} contribution to $\th$, i.e.
$\th_{\rm ind} = - \ch_{{\cal O}_{\rm CP}}(0)/\ch(0)$,  is
linearly related to the coefficient of the $CP$-odd operator ${\cal O}_{\rm CP}$ generating 
$\ch_{{\cal O}_{\rm CP}}(0)$. These effects therefore need to be taken into account in computing the 
observable consequences of $CP$-odd sources in axion scenarios, and will be important for us later on.

Before moving on, it is worth recalling that, were it realized, the simplest solution to the 
strong $CP$ problem would fall into the class we are discussing, namely the possibility that 
$m_u=0$ in the Standard Model Lagrangian normalized at a high scale $M$, or more generically, 
det$(Y_u(M))=0$. In this situation, the Lagrangian already possesses the appropriate chiral symmetry without the
addition of extra fields and, as we have discussed, $\th(M)$ then becomes unphysical. Since the identification
of light quark masses is indirect, using meson and baryon spectra and chiral perturbation theory, the 
possibility that $m_u=0$ has been debated at length in the literature \cite{KapMan,Nir}, but is
strongly disfavoured by conventional chiral perturbation theory analysis, with recent results implying 
$m_u/m_d = 0.553\pm 0.043$ \cite{GL}, and this conclusion is beginning to be backed up by unquenched  
(but chirally extrapolated) lattice simulations which suggest similar values, $m_u/m_d = 0.43 \pm 0.1$ \cite{milc}.

\subsubsection{Engineering $\bar\theta\simeq 0$ } 

Another way to approach the strong $CP$ problem is to assume that either $P$ or $CP$ or 
both are exact symmetries of nature at some high-energy scale. Then one can declare 
that $\theta G\tilde G$ be zero at this high scale as a result of symmetry. 
Of course, to account for the parity- and $CP$-violation observed in the 
SM, one has to assume that these symmetries are spontaneously broken at a particular scale $\La_{P(CP)}$.

The model building problem that this sets up -- one which has been made particularly manifest by
the consistency of the recently observed $CP$-violation in $B$-meson decays with the 
KM mechanism -- is that one needs to ensure that the subsequent corrections to $\th$ are small, 
while still allowing for an order one KM phase. Symmetry breaking at $\La_{P(CP)}$ may generate the 
$\th$-term at tree level through
e.g. imaginary corrections to the quark mass matrices $M_u$ and $M_d$,
\ba
 \bar{\th} \sim \mbox{Arg Det}(M_u M_d)+\cdots \label{thind}
\ea
where such corrections could affect either the Yukawa couplings $Y_{u(d)}$ or the Higgs vacuum
expectation values, $v_{u(d)}$ (in the SM $v_u = v_d^*$), while the ellipsis denotes the 
phases of other coloured fermions.
For comparison, the SM CKM-type phase (in basis-invariant form) is \cite{jarlskog}
\ba
 \th_{\rm KM} \sim \mbox{Arg Det}\left[M_uM^\dagger_u,M_dM^\dagger_d\right],\label{CKMphase}
\ea
and one is then led to consider models for flavour in which the second phase (\ref{CKMphase})
can be large, as is required, while the first (\ref{thind}) vanishes, or is at least 
highly suppressed. 

One class of models uses exact parity symmetry at some high-energy scale which
implies $L\leftrightarrow R$ reflection symmetry 
in the Yukawa sector and thus hermitian Yukawa matrices which do not contribute to $\bar\th$ \cite{MohSen}.
However, this necessitates the extension of the SM gauge group to incorporate $SU(2)_R$, and the 
reality of $v_{u(d)}$ comes as an additional constraint on the model which can be achieved {\em e.g.} in its 
supersymmetric versions \cite{Kuchi,MohRa}. One can instead just demand that $CP$ be an exact symmetry
at high scales, which is then broken spontaneously and $CP$-violation enters via complex vacuum expectation 
values of additional scalar fields. Models of this type can be constructed in 
which the mass matrices are complex,
but have a real determinant \cite{Nelson,Br}, although often it is difficult to obtain a sufficiently
large CKM phase. An interesting recent suggestion for getting round this problem is to use low-scale 
supersymmetry
breaking \cite{HilSm} (see also earlier ideas \cite{Holdom,Ratt}), while $CP$ is broken spontaneously
at a much higher scale where SUSY is still exact. Strong interactions in the $CP$-breaking sector can
then generate a large CKM phase, while a SUSY nonrenormalization theorem ensures that $\bar\th$ is
not generated until the much lower scale where SUSY is broken.

All models of the type discussed above, that attempt to solve the strong $CP$ problem by 
postulating exact parity or $CP$ at high scales, have to cope with the very tight bound on 
$\bar\theta$. Indeed, it is not enough to obtain $\bar\theta=0$ at tree level, as loop effects at 
and below $\Lambda_{P(CP)}$ can lead to a  substantial renormalization
of the $\th$-term (see, {\rm e.g.} \cite{KD,Pospelov}). If the effective theory reduces to the SM
below the scale $\Lambda_{P(CP)}$, the residual {\it low-scale} corrections to the $\th$-term
can only come via the Kobayashi-Maskawa phase and the resulting 
value for $\bar\theta(\delta_{KM})$ is small. However, this does not guarantee that the 
threshold corrections at $\Lambda_{P(CP)}$ 
are also small, as they will depend on different sources of $CP$-violation 
and do not have to decouple in the limit of large $\Lambda_{P(CP)}$. 
Such corrections are necessarily model-dependent. However, if  
the underlying theory is supersymmetric at the scale $\Lambda_{P(CP)}$ and 
the breaking of supersymmetry occurs at a lower scale 
$\Lambda_{SUSY}$, one expects the corrections to $\bar\theta$ to be suppressed by 
power(s) of the small ratio $\Lambda_{SUSY}/\Lambda_{P(CP)}$ \cite{HilSm}.

To summarize this section, we comment that the way the strong $CP$ problem is 
resolved affects  the issue of how large additional non-CKM $CP$ violating sources can be.  
The axion solution, as well as $m_u=0$, generically allows for the presence of 
{\em arbitrarily} large $CP$ violating sources above a certain energy scale. 
This scale is determined by comparison of higher-dimension $CP$-odd operators
(i.e. dim$\geq 5$) induced by these sources with the current EDM constraints. 
On the contrary, models using a discrete symmetry solution to the strong $CP$ problem
usually have tight restrictions on the amount of additional $CP$-violation even at 
higher scales in order to avoid potentially dangerous contributions to the $\th$-term.

\section{QCD Calculation of EDMs}

Having discussed the $\th$-term in detail, we now take a more general approach and consider all the
relevant operators up to dimension six in the $CP$-odd Lagrangian (\ref{456}), and move to the
next level in energy scale in Fig.~1. To proceed, we need to determine the dependence of the 
nucleon EDMs, pion-nucleon couplings, etc., on these quark-gluon Wilson coefficients normalized 
at 1 GeV, i.e.
\ba
 &&d_n = d_n(\bar\theta, d_i, \tilde d_i, w, C_{ij}),\nonumber\\
 &&\bar g_{\pi NN} = \bar g_{\pi NN}(\bar \theta, \tilde d_q, w, C_{ij}),
\label{goal}
\ea
The systematic project of deducing this dependence was first initiated some 20 years ago by 
Khriplovich and his collaborators, and is clearly a nontrivial task as it involves nonperturbative 
QCD physics. It is nonetheless crucial in terms of extracting
constraints, and in particular one would like to do much better than order of magnitude estimates so
that the different dependencies of the observable EDMs may best be utilized in constraining models
for new physics. 

It is this problem that we will turn to next. In order to be concrete, we will limit our discussion to the
nucleon EDMs and pion-nucleon couplings. The electron nucleon couplings, of which $C_S$ plays
the most important role for the EDM of paramagnetic atoms, receive contributions
from the semi-leptonic four-fermion couplings $C_{qe}$ in (\ref{manydim6}), which
may be determined straightforwardly using low-energy theorems for the matrix elements of
quark bilinears in the nucleon (See {\em e.g.} Refs. \cite{KBM}).

Before we delve into some of the details of these calculations, it is worth outlining a checklist of
attributes against which we can compare the various techniques available for these calculations. We list
below several features that such techniques would {\it ideally} possess:

\begin{itemize}
\item {\it Chiral invariance}, including the 
relevant anomalous Ward identities, provides a very strong contraint
on the manner in which $CP$-odd sources may lead to physical observables in the QCD sector. As an example, 
distributng $\bar \theta$ arbitrarily between $G\tilde G$ and $\bar q i \gamma_5 q$ cannot 
alter the prediction for $d_n(\bar \theta)$, and the answer must also depend on the correct 
combination of quark masses, namely the reduced mass $m_*$. These symmetry constraints are therefore very
powerful, and allow a consistency check of the QCD estimates. Calculations are therefore more transparent
if these constraints can be ``built in''. 

\item In addition to these chrial properties, the need to deal first of all with the tuning of $\bar\th$ means
that ideally the procedure should also corrrectly account for additional contributions generated under
PQ relaxation. As argued in the previous section, the presence of $CP$-odd 
operators can shift the position of the axion expectation value, leading to a new class of 
contributions, $d_n(\theta_{\rm ind})$. More specifically, this is the case for the CEDM sources, the
presence of which implies that $\bar{\th}$ must be substituted not by zero but 
by $\theta_{\rm ind}$ given by \cite{BUP}:
\ba
 \th_{\rm ind} = -\frac{m_0^2}{2}\sum_{q=u,d,s}
     \frac{\tilde d_q}{m_q},
     \label{theta_ind}
\ea
independently of the specific details of the axion mechanism. 
Here $m_0^2$ determines the strength of the following mixed quark-gluon condensate,
\ba
 g_s\langle \ov{q} G\si q \rangle \equiv  m_0^2\qq,  \label{m021}
\ea
Thus PQ symmetry may lead to additional vacuum contributions to the EDM. 

\item 
In order to make consistent use of the EDM constraints on fundamental $CP$-odd phases, 
it would be desirable to have the {\em same method} available for obtaining 
estimates of  $d_n$ and $\bar g_{\pi NN}$ in terms of the relevant Wilson coefficients.
Since different techniques have different sources of errors,  use of the same method may allow 
a reduction in the uncertainty between the relative coefficients which, given a suite of different constraints,
is ultimately more important then the overall uncertainty. 

\item 
Ideally, the method used should allow for a systematic {\it estimate of the precision} of the
relevant QCD matrix elements, i.e. within a framework allowing for a 
treatment of higher order subleading corrections.

\item The generic dependence of $d_q$ on $m_q$ poses an additional challenge for obtaining precise
results, through  the poorly known values of the light quark masses, and their strong dependence on
the QCD normalization scale. This uncertainty can be ameliorated given a method which generates
answers which depend only on scale invariant combinations such as {\em i.e.} 
$(m_u+m_d)\qq = -f_\pi^2m_\pi^2$ or $m_u/m_d$.

\item 
For EDMs, the contribution of operators that are not suppressed by the light quark masses, 
$m_u$ and $m_d$, is of considerable phenomenological interest. An ideal method would lead to a 
quantitative prediction for whether $\tilde d_s$ or $w$ can compete with the contributions of 
light quark EDMs and CEDMs. A well-known example in the $0^+$ channel suggests that this 
may indeed happen: i.e. $\langle N|m_s\bar ss |N\rangle >  \langle N|
m_q\bar qq |N\rangle$, where $q=u,d$.

\end{itemize}

Since this is a nonperturbative QCD problem, the tools at our disposal are limited. 
Ultimately, the lattice may provide the most systematic 
treatment, but for the moment we are limited to various appoximate methods and none that
are currently available can satisfy all of the demands listed above, although we will argue that 
combining QCD sum-rules with chiral techniques can satisfy most of them. 
While one can make use of various models of the infrared regime of
QCD, we prefer here to limit our discussion to three (essentially) model-independent approaches, which
vary both in their level of QCD input, and in genericity as regards the calculations to which 
they may be applied.

However, we will first recall what is perhaps the most widely used approach for estimating the
contribution of quark EDMs to the EDM of the neutron. This is the use of the SU(6) quark model,
wherein one associates a nonrelativistic wavefunction to the neutron which includes three constituent
quarks and allows for the two spin states of each. Obtaining the contribution of quark EDMs to $d_n$ then amounts
to evaluating the relevant Clebsch-Gordan coefficients and one finds,
\ba
 d_n(d_q)^{\rm QM} = \frac{1}{3}(4d_d-d_u).
\ea
Although one may raise many questions regarding the reliability, and expected precision, of 
this result, we will emphasize 
here only the significant disadvantage that this approach cannot be used for a wider class of CP-odd
sources, relevant to the generation of $d_n$ and $\bar{g}_{\pi NN}$.

\subsection{Naive dimensional analysis}

Although historically not the first, conceptually the simplest approach is a form of QCD power-counting
which goes under the rather unassuming name of ``naive dimensional analysis'' (NDA) \cite{GM}.
This is a scheme for estimating the size of some induced operator by matching loop corrections to the 
tree level term at the specific scale where the interactions become strong. In practice, one uses
a dimensionful scale $\La_{\rm had}\sim 4\pi f_\pi$ characteristic of chiral symmetry breaking, and a 
dimensionless coupling $\La_{\rm had}/f_\pi$ to parametrize the coefficients. The claim is that, to within
an order of magnitude,
the dimensionless ``reduced coupling'' of an operator below the scale $\La_{\rm had}$ is given 
by the product of the reduced couplings of the operators in the effective Lagrangian 
above $\La_{\rm had}$ which are required to generate it. The reduced couplings are determined by demanding 
that loop corrections match the tree level terms, and for the 
coefficient $c_{\cal O}$ of an operator ${\cal O}$ of dimension $D$, containing $N$ fields, is 
given by $(4\pi)^{2-N}\La_{\rm had}^{D-4} c_{\cal O}$. A crucial, and often rather delicate, point 
is the precise scale at which one should perform this
matching. Within the quark sector, the identification of this scale with $4\pi f_\pi$ often
seems to work quite well. However, for gluonic operators, the implied matching occurs at a very low
scale where $g_s$ is very large, up to $g_s\sim 4\pi$, and NDA has proved more problematic in this sector.

To illustrate this approach, let us consider the neutron EDM induced by $\th$, in this case realized
as an overall phase $\th_q$ of the quark mass matrix, and also the EDM and CEDM of a light quark. 
The dimension five neutron EDM operator has reduced coupling $d_n \La_{\rm had}/(4\pi)$. Above the scale 
$\La_{\rm had}$  we need the reduced couplings of the electromagnetic coupling of the quark, $e/(4\pi)$, and
the $CP$-odd quark mass term, $\th_q m_q/\La_{\rm had}$. Thus we find,
\ba
 d_n(\th_q,\mu) \sim e \th_q(\mu) \frac{m_q(\mu)}{\La_{\rm had}^2},
\ea
where the $\mu$-dependence reflects the choice of matching scale. To obtain a similar estimate
for the contribution of a light quark EDM, we note simply that it has a reduced coupling given by 
$d_d \La_{\rm had}/(4\pi)$ and thus
\ba
 d_n(d_q,\mu) \sim d_q(\mu),
\ea
which can be contrasted with the quark model estimate above. The contribution of the quark CEDM is
similar, but one needs in addition the reduced electromagnetic coupling of the quark, $e/(4\pi)$, so
that 
\ba
 d_n(\tilde{d}_q,\mu) = \frac{e g_s(\mu)}{4\pi}\tilde{d}_q'(\mu),
\ea 
where we have redefined the CEDM operator so that $\tilde{d}_q = g_s\tilde{d}_q'$. This makes the
factor $g_s$ explicit, which seems crucial to the success of NDA for gluonic operators as the matching
needs to be performed at a large value of $g_s$, e.g. $g_s\sim 4\pi$ as noted above.

These examples indicate on one hand the simplicity of this approach and also its general
applicability, but also the fact that it does not easily allow one to combine different contributions
into a single result for the neutron EDM. In particular, these estimates have uncertain signs and
thus can only be used independently with an assumption that the physics which generates them does not
introduce any correlations. This will not generically be the case.

\subsection{Chiral techniques}

Historically, the first model-independent calculation of the neutron EDM  \cite{CDVW} 
made use of chiral techniques to isolate an infrared log-divergent contribution in the chiral limit (for an 
earlier bag model estimate see \cite{baluni}).
This was one of the landmark calculations which made the strong $CP$ problem, and indeed the magnitude of the
required tuning of $\th$, quite manifest.

The basic observation was that, given a $CP$-odd pion-nucleon coupling $\bar{g}_{\pi NN}$, one 
could generate a contribution to the neutron EDM via a $\pi^-$-loop (see Fig.~2) which was 
infarared divergent in the chiral limit.  In reality this log-divergence is cut off by 
the finite pion mass, and one obtains,
\ba
 d_n^{\ch {\rm log}} = \frac{e}{4\pi^2 M_n} g_{\pi NN}\bar{g}_{\pi NN}^{(0)} \ln \frac{\La}{m_\pi},
  \label{clog}
\ea
where $\La$ is the relevant UV cutoff, i.e. $\La=m_\rho$ or $M_n$. One 
can argue that such a contribution cannot be systematically 
cancelled by other, infrared finite, pieces and thus the bound one obtains on $\bar{g}^{(0)}_{\pi NN}$
in this way is reliable in real-world QCD.

\begin{figure}[t]
 \centerline{%
   \includegraphics[width=8cm]{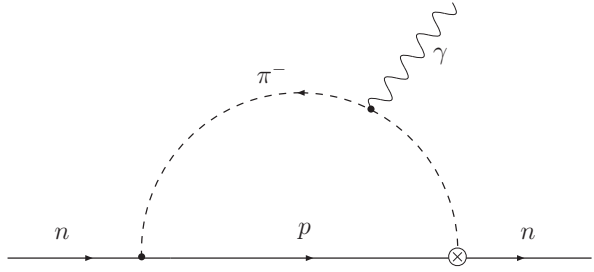}%
         }
\vspace{0.1in}
 \caption{Chirally enhanced contribution to the neutron EDM..}
\end{figure}

This reduces the problem to one of computing the relevant $CP$-odd pion-nucleon couplings. For a given
$CP$-odd source ${\cal O}_{CP}$, we have 
\ba
 \langle N \pi^a | {\cal O}_{CP} |N^\prime\rangle =
\fr{i}{f_\pi}\langle N|[ {\cal O}_{CP}, J^a_{05} ] |N^\prime\rangle + {\rm rescattering},
\label{start}
\ea
justified by the small $t$-channel pion momentum. The possible rescattering corrections will be
discussed below. If we now specialize to the $\th$-term, as
in \cite{CDVW}, with ${\cal O}_{CP} = -\th_q m_* \sum_f \bar{q}_f i\ga_5 q_f$ then the 
commutator reduces to the triplet nucleon sigma term, and we find
\ba
 \bar{g}^{(0)}_{\pi NN}(\th_q) = \frac{\th_q m_*}{f_\pi} \langle p | \bar{q}\ta^3 q |p\rangle
  \left( 1- \frac{m_\pi^2}{m_\et^2}\right). \label{gpiNNscalar}
\ea
One can then determine $\langle N | \bar{q}\ta^a q |N\rangle$ from lattice calculations or, as
was done in \cite{CDVW}, by using global SU(3) symmetry to relate it to measured splittings in the
baryon octet.

The final factor on the right hand side of (\ref{gpiNNscalar})
reflects the vanishing of the result in the limit that
the chiral anomaly switches off and $\et$ (or $\et'$ in the three-flavour case) is a genuine
Goldtone mode. This factor is numerically close to one and was ignored in \cite{CDVW}. It arises
because in (\ref{start}) we should also take into account the fact that the $CP$-odd mass term can produce
$\et$ from the vacuum and thus, in addition to the PCAC commutator, there are rescattering graphs
with $\et$ produced from the vacuum and then coupling to the nucleon, and the soft pion radiated 
via the $CP$-even pion-nucleon coupling \cite{pr3}. 

Although this technique is not universally applicable, one can also contemplate 
computing the contribution of certain other sources, e.g. the quark CEDMs.
Using the same PCAC-type reduction of the
pion in $\langle N \pi^a | {\cal O}_{CP} |N^\prime\rangle$ as in (\ref{start}), 
one can reduce the calculation of $\bar g_{\pi NN}$ to the matrix elements 
of dimension five $CP$-even operators. In doing this, one has to 
take into account a subtlety for CEDM sources, first
pointed out in \cite{FOPR,pr3,pr4}, namely that a second class of contributions, the
pion-pole diagrams (Fig.~3b), now contribute at the same order in chiral perturbation theory. 
In an alternative but physically equivalent approach, one can perform a chiral rotation in 
the Lagrangian to set $\langle 0| {\cal L}_{\;\,\slash \!\!\!\!\!CP}|\pi^0\rangle=0$, thus making this additional
source of $CP$-violation explicit at the level of the Lagrangian \cite{pospelov01}.
\begin{figure}[t]
 \centerline{%
   \includegraphics[width=11cm]{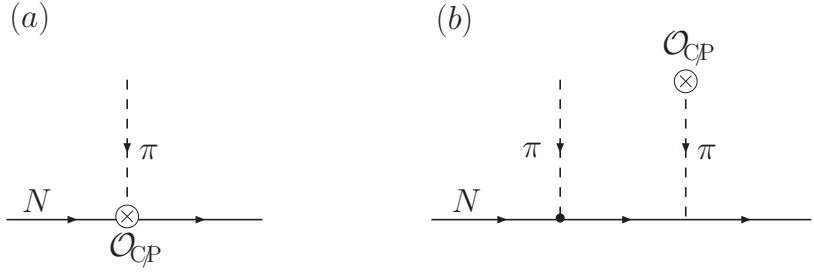}%
         }
 \caption{Two classes of diagrams contributing to the CP-odd pion-nucleon
coupling constant.}
\end{figure}

After this consistent PCAC reduction of the pion, the intermediate result 
for $\bar{g}_{\pi NN}$ takes the following form,
\begin{eqnarray}
&&\fr{1}{2f_\pi}\langle N | \tilde d_u(\bar u g_sG\si u - m_0^2
\bar uu) - \tilde d_d (\dGd - m_0^2\dd )|N\rangle  \nonumber\\ &&+
\fr{m_*}{f_\pi}(\bar\theta -\theta_{\rm ind})
\langle N | \bar{q}\ta^3 q|N\rangle.
\label{chiral_g}
\end{eqnarray}
The second line in this expression contains the same matrix element as
(\ref{gpiNNscalar}), and vanishes when  PQ symmetry sets the axion minimum
to $\theta_{\rm ind}$ (\ref{theta_ind}). The remaining terms are proportional
to the following combination,
\ba
 \langle N| \ov q g_s G\si q - m_0^2 \ov{q} q  |N\rangle,
 \label{thisiswhatweneed}
\ea 
which unfortunately {\em cannot} be estimated using chiral techniques, and requires
genuine QCD input. A naive vacuum saturation hypothesis in (\ref{thisiswhatweneed}) 
leads to the vanishing of this expression. This is a rather fundamental problem which limits
the precision of various approaches, {\it e.g.} those based on the use of low energy theorems to 
estimate (\ref{thisiswhatweneed}) \cite{KKY,KKZ}, to obtain the dependence of $\bar{g}_{\pi NN}$ on the CEDMs. 

This limited applicability is one problem that currently afflicts the chiral approach. A more
profound issue is that the terms enhanced by the chiral log, while conceptually distinct, are not necessarily
numerically dominant. Indeed there are infrared finite corrections to (\ref{clog}) which, while
clearly subleading for $m_{\pi}\rightarrow 0$, are not obviously so in the physical regime. This
dependence on threshold corrections has been observed to 
provide a considerable source of uncertainty \cite{PdR} (see also \cite{skyrme}).

\subsection{QCD sum-rules techniques}

An alternative to considering the chiral regime directly, is to first start at high energies,
making use of the operator product expansion, and attempt to construct QCD sum rules \cite{svzsr} for the
nucleon EDMs, or the $CP$-odd pion nucleon couplings. This approach in principle allows for
a systematic treatment of all the sources, and is motivated in part by the success of such approaches
to the calculation of baryon masses \cite{ioffe81} and magnetic moments \cite{is}. 
For a recent review of some aspects of the application of QCD sum rules to nucleons, 
see {\em e.g.} Ref. \cite{leinweber}.

The basic idea is familiar from other sum-rules applications. One considers the two-point correlator
of currents, $\et_N(x)$, with quantum numbers of the nucleon in question 
in a background with nonzero CP-odd sources, an electromagnetic field $F_{\mu\nu}$, and also a 
soft pion field $\pi^a$, 
\ba
 \Pi(Q^2) & = & i\int d^4x e^{ip\cdot x}
    \langle 0|T\{\et_N(x)\ov{\et}_N(0)\}|0\rangle_{\cp,F,\pi},
     \label{pi}
\ea
where $Q^2=-p^2$, with $p$ the current momentum. It is implicit here that the soft pion field admits
PCAC reduction, and then in the case of CEDM sources corresponds to an external field coupled to 
the operator $\bar q g_sG\si q -m_0^2 \bar qq$, as in (\ref{chiral_g}-\ref{thisiswhatweneed}). 

One then computes the correlator at large $Q^2$
using the operator product expansion (OPE), generalised to incorporate condensates of the fields, and then
matches this to a phenomenological parametrization corresponding to an expansion of the nucleon
propagator to linear order in the background field and $CP$-odd sources, and 
corresponding higher excited states in the relevant channel. 
In practice, one makes use of a Borel transform to suppress the contribution
of excited states, and then checks for a stability domain in $Q^2$, or rather the 
corresponding Borel mass $M$, where the two asymptotics may be matched. 

Let us discuss this procedure in a little more detail.
For the neutron, there are two currents with lowest dimension that are commonly used in
QCD sum-rules and lattice calculations,
\ba
 \et_1 &=& 2\ep_{abc}(d_a^TC\ga_5u_b)d_c, \nonumber\\
 \et_2 &=&  2\ep_{abc}(d_a^TCu_b)\ga_5d_c, \label{j1j2}
\ea
of which the second, $\et_2$, vanishes in the nonrelativistic limit,
and lattice simulations have shown that $\et_1$ indeed provides the 
dominant projection onto the neutron state
\cite{leinweber,chung82}. In the truncated OPE expansion, an
admixture of $\et_2$ can nonetheless be used to optimise convergence, and thus
it is natural to parametrize the current in the form $\et_n=\et_1+\beta \et_2$
introducing an unphysical parameter $\beta$. The truncated OPE will then inherit
a dependence on this parameter, which can then be fixed to improve convergence
once the sum-rule has been constructed for a given physical quantity. 

The correlator (\ref{pi}) exhibits various Lorentz structures (LS) in its 
OPE and, in selecting one to consider, one needs to be aware that in a
$CP$-violating background the coupling of the current to the neutron state,
described by a spinor $v$, is not invariant under chiral rotations, i.e. 
$\langle 0 | \et_n  |0\rangle = \la e^{i\al\ga_5/2} v$. It turns out that
of the terms contributing to the neutron EDM, there is a unique structure which is
invariant under chiral rotations, namely LS=$\{F\si\ga_5,\psl\}$, thus this is
the natural quantity on which to focus in constructing a sum rule for the EDM (for alternatives,
see \cite{kw,henley}).
Correspondingly, LS=$\psl$ is the relevant chiral-invariant structure 
for the $\bar g_{\pi NN}$ sum rule. 

In constructing a phenomenological model for the current correlator, it is apparent that
in expanding to linear order in the external field we are effectively considering a three-point function.
It is then not particularly useful to work with the spectral density, as is standard for two-point functions,
and the conventional approach is to parametrize the correlator itself, i.e. 
$\Pi^{\rm phen} ={\rm LS}f(p^2)+\cdots$. The function $f(p^2)$ will in general have an expansion 
in double and single pole terms, and then a continuum modelling the transitions between excited states.
A Borel transform can be applied to suppress the contribution of excited states.
However, a well-known complication \cite{is} of baryon sum rules in
external fields is that the single pole terms, corresponding to 
transitions between the neutron and excited states, are not
exponentially suppressed by the Borel transform and thus provide 
the leading contribution from the excited states, with a coefficient 
which is not sign definite. This must then be treated as a
phenomenological parameter to be determined from the sum rules
themselves. In this approximation, we then find \cite{pr2,pr3,pr5},
\ba
 && \Pi^{\rm phen}_{(d)} =\frac{i}{2}\{F\si\ga_5,\psl \}\left(
\fr{\lambda^2d_nm_n}{(p^2-m_n^2)^2} +
\fr{A}{p^2-m_n^2}+\cdots\right), 
          \label{phenfull}\\
          && \Pi^{\rm phen}_{(\bar{g})} = 2\psl \left(
\fr{\lambda^2\bar g_{\pi NN} m_n}{(p^2-m_n^2)^2} +
\fr{A'}{p^2-m_n^2}+\cdots\right),
\ea
where the constants $A,A'$ parametrise the single-pole contributions.
One can then go further and construct a full continuum model to match the
high-$Q^2$ asymptotics, but as discussed below this refinement has minimal impact in comparison to the
single pole terms $A$ and $A'$.
We now turn to the calculation of the OPE for $d_n$ and $\bar g_{\pi NN}$.

\begin{figure}[t]
 \centerline{%
   \includegraphics[width=6.5cm]{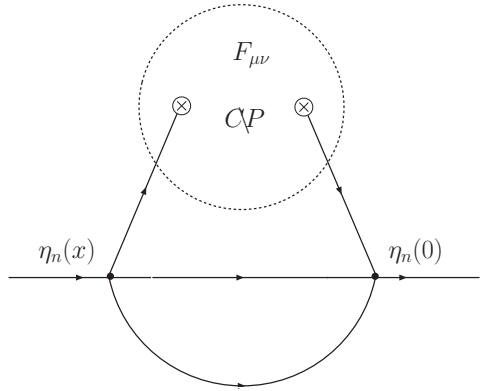}%
         }
 \caption{A leading contribution to the neutron EDM within QCD sum rules. Sensitivity to the $CP$-violating 
source enters through the two soft quark lines which lead to a dependence on the chiral condensate.}
\end{figure}

\bigskip
$\bullet$ {\it Nucleon EDM calculations}
\bigskip

The OPE for $d_n$ is conveniently constructed in practice by first computing the generalized
quark propagator, expanded in the presence of the background field, the
CP-odd sources, and also the vacuum condensates. One then computes the relevant contractions in (\ref{pi})
to obtain the OPE to the appropriate order. 
Although it would take us too far afield to describe this procedure in detail, we can exhibit some of the
dominant physics by looking at just one class of diagrams which arise in evaluating the OPE for
(\ref{pi}). In particular, in Fig.~4 two of the quarks in the nucleon current propagate without 
interference, carrying the large current momentum, while the third is taken to be soft and
so induces a dependence on the appropriate chiral quark condensates. We may then make use of standard
arguments \cite{svzcp} used to determine $E(\theta)$ (\ref{Energy}), which utilize the 
anomaly in the axial current
and the fact that $m_{\et}\gg m_{\pi}$, to determine the dependence of these
condensates on the $CP$-odd source. 

Let us consider the contribution of $\bar\th$.
For additional control over the chiral transformation properties of the answer, we split $\bar\th$ into several 
terms in the Lagrangian,
\ba
{\cal L}_{\rm dim=4}=\frac{g_s^2}{32\pi^{2}}\ \theta_G
G^{a}_{\mu\nu} \widetilde{G}^{\mu\nu , a} - \sum_{i=u,d,s} m_i \th_i \ov{\ps}_i i\ga_5 \ps_i
\ea
The diagram in Fig.~4 then leads to a dependence on
\ba
 m_q \langle 0| \bar{q} \si_{\mu\nu}\ga_5 q | 0\rangle_{\th,F} &=& 
   im_* \th_G \langle 0| \bar{q} \si_{\mu\nu} q | 0\rangle_{F} + {\cal O}(m_*^2) \nonumber\\
 &=& i\ch_q \th_G m_* F_{\mu\nu} \langle  \ov{q}q\rangle + {\cal O}(m_*^2).
 \label{chi-cp}
\ea
In the first equality the dependence on $\th_G$ has been determined as in the standard computation of
$E(\theta)$ in (\ref{Energy}) \cite{svzcp}, where the terms of ${\cal O}(m_*^2)$ are subleading only
because there is no U(1)-axial Goldstone mode, i.e. $m_{\et}\gg m_{\pi}$. 
The dependence of (\ref{chi-cp}) on $\th_G$ rather than $\bar \th$ means that there are 
additional contributions from $\theta_q$ that have to be taken into account, at this order in the OPE, in order
to restore the dependence on the physical $\bar\theta$ combination. In fact, it turns out that the 
procedure is generally more complicated and one must also incorporate the the mixing of the currents
$\et_1$ and $\et_2$ (\ref{j1j2}) with their chirally-rotated (or $CP$-conjugate) counterparts \cite{pr2,pr3}
in order to restore the dependence on $\bar\theta$ and exclude unphysical parameters such as $\th_G-\th_q$.

In the second line of (\ref{chi-cp}) we have introduced the so-called  electromagnetic susceptibility of the 
quark vacuum, $\ch_q$, given by 
\ba
 \langle \ov{q}\si_{\mu\nu}q\rangle_F =  \ch_q F_{\mu\nu}
           \langle  \ov{q}q\rangle, \label{chidef}
\ea
and chosen for simplicity to be favour independent, $\ch_q=\ch e_q$ \cite{is}.
In fact, $\chi_q$ is one among a whole series of mixed quark-gluon condensates that have to 
be taken into account in the calculation of the OPE for $d_n$.
However, it turns out that numerically $\chi$ is rather large, $\ch = - (5-9)$ GeV$^{-2}$ 
\cite{chival,vainshtein}, 
and results in the diagram of Fig.~4 being numerically very important. The remaining condensates
are numerically quite
small in comparison, and we will neglect them in what follows, 
referring the reader to Refs.~\cite{pr1,pr2,pr3,pr4,pr5} 
for more of the details involved in these calculations.

The leading order and next-to-leading order contributions to the
OPE induced by the $CP$-odd operators of dimension four and five are given by
\ba
 \Pi^{\rm OPE}(Q^2) & = & -\frac{i}{64\pi^2}\qq\{F\si\ga_5,\psl\} \nonumber\\
 &&\;\;\;\;\;\times \left( \ln(-p^2)[\pi_{\rm LO}^{(\ch)}
     +\pi_{\rm LO}^{(q)}] -\frac{4}{p^2}\ln\left(-\frac{\La_{IR}^2}{p^2}\right)[\pi_{\rm NLO}]\right).
\label{ope}
\ea
At leading order the quark EDMs induce $\pi_{\rm LO}^{(q)}$
\ba
 \pi_{\rm LO}^{(q)} &=& d_d\left[10+6\beta^2\right]-d_u\left[3+2\beta-\beta^2\right],
 \label{qedm}
\ea
while the $\th$-term, and the CEDMs, are responsible for $\pi_{\rm LO}^{(\ch)}$,
\ba
 \pi_{\rm LO}^{(\ch)} &=& 4(1+\beta)^2\ch_dm_dP_d-(1+\beta)^2\ch_um_uP_u \nonumber\\
          &&\;\;\;\;\;\;\;\;\;\;\;\;\;\;\;\;\;\;\;\;\;\;\;\;
     +2(1-\beta^2)m_*(\ch_u+\ch_d)(P_u-P_d), \label{chi}
\ea
where $P_q=\th_q + \frac{\qqg_{\cp}}{i\qq}$ contains the dependence on the $CP$-odd sources.

The next-to-leading order terms in (\ref{ope}), $\pi_{\rm NLO}$, are associated with 
a class of diagrams in which one of the propagators exhibits a logarithmic infrared divergence as can be 
seen in (\ref{ope}). The magnitude of these terms is clearly ambiguous through the 
logarithmic dependence on the cutoff $\La_{IR}$, although in practice
with a cutoff $\La_{IR}\sim \La_{QCD}$ the logarithms are not 
particularly large in the momentum regime for which the resulting
sum rules are optimal. We will not need to exhibit them explicitly here, but it is important 
that for all of these terms $\pi_{\rm NLO}\propto (1-\beta)$, which differs from the $(1+\beta)^2$ 
dependence of many of the dominant leading order terms.

The truncated OPE in (\ref{ope}) necessarily depends on the 
unphysical parameter $\beta$, which can therefore be chosen in such a way
as to optimise the convergence of the expansion. Such optimization problems
arise in many areas of physics \cite{stevenson}, and in the present case with
only two terms of the series in hand, the most practical approach is to use
``fastest apparent convergence'' (FAC), which involves choosing $\beta$
to set the highest known term in the series to zero.  Historically, Ioffe
\cite{ioffe83} introduced an FAC-like criterion in analysis of the
mass-sum rule, which has proved quite successful. For the $CP$-even sector 
this involves the choice $\beta=-1$ to cancel the subleading terms. We shall
also follow this approach as it has the added advantage of canceling the ambiguous
infrared logarithmic terms. As first discussed in \cite{pr2,pr3}, for 
the $CP$-odd sector this involves the choice $\beta=1$. The difference
in $\beta$, as compared with the choice in the $CP$-even sector, 
is not surprising as $\beta$ is unphysical and there is then no reason to expect optimal choices
to be the same for different physical observables.
  
A remarkable feature of the choice $\beta=1$ becomes apparent if we now
rewrite the OPE expression for $\beta=1$. All the subleading 
logarithmic terms in (\ref{ope}) are now cancelled, while the 
leading order terms adopt the elegant form,
\ba
 \pi^{(\ch)}_{\beta=1} & = & 4\left[4\ch_dm_dP_d-\ch_um_uP_u\right] \nonumber\\
 \pi^q_{\beta=1} & = & 4\left[4d_d-d_u\right] .
\label{pibeta1}
\ea
It is remarkable that the contribution of
the $u$ quark in each term is precisely $-1/4$ that of the $d$ quark,
which is the combination suggested by the SU(6) quark model! 
One might suspect that this is due to the minimal valence quark content of
the current, but the fact that this structure arises only
for the choice $\beta=1$ is rather surprising given that only the 
$\et_1$ current survives in the nonrelativistic limit.

One of the most important aspects of the whole calculation is 
the consistent treatment of the $CP$-odd vacuum condensates
entering via the $CP$-odd sources $P_q$. This 
calculation can be done relatively easily with the use of chiral techniques, and 
at leading order in $m_{\pi(\eta)}^2/m_{\eta'}^2$.
A useful constraint on this calculation is that the anomalous chiral 
Ward identity must be respected for each quark flavour.
This provides a useful check if we consider artificially 
decoupling the s-quark, while at the same time sending either 
$m_u$ or $m_d$ to zero. In this regime, the dependence on $\bar{\th}$ 
in particular must vanish, which fixes the remaining quark mass 
dependence in terms of $m_*$. The final result for the $P_u$ and $P_d$ 
sources, in the limit of $m_{u(d)}\ll m_s$, reads
\ba
 m_{u(d)}P_{u(d)} &=& m_*\bar\th -
 \frac{m_*m_0^2}{2}\left(\frac{\tilde{d}_{u(d)}-\tilde{d}_{d(u)}}{m_{d(u)}}
 -\frac{\tilde{d}_s}{m_s}\right) \label{Pq}, 
\ea
which respects the anomalous chiral Ward identity. If $\bar\theta$ is removed by PQ
symmetry, i.e. substituted by $\theta_{\rm ind}$ (\ref{theta_ind}),
Eq. (\ref{Pq}) simplifies even further,
\ba
m_q P_{q}^{\rm PQ} = - \fr{m_0^2 \tilde d_q}{2}.
\ea

Putting all the ingredients together, after a Borel transform of (\ref{ope}) and (\ref{phenfull}), and
using $\beta=1$ as discussed earlier, we obtain the sum rule,
\ba
 \la^2 m_nd_n+AM^2 &=&
-\frac{M^4}{32\pi^2}e^{m_n^2/M^2}\qq
        \left[\pi^{(\ch)}_{\beta=1}
+\pi^q_{\beta=1}
 \right] +O(M^0),
   \label{sumrule}
\ea  
where $M$ is the Borel mass, and the relevant contributions are given in (\ref{pibeta1}).
Clearly, the presence of three parameters $d_n$, $\la$ and $A$ in (\ref{sumrule})
necessitates the use of additional sum rules, and the coupling $\la$ 
is conveniently obtained from the well-known sum rules for 
the two-point correlation function of the nucleon currents in the $CP$ even sector 
(see e.g. \cite{leinweber} for a review). 

Rather than reviewing the full analysis, let us consider a simple
estimate obtained from the leading order terms in the OPE of
(\ref{sumrule}) a la Ioffe's derivation of the nucleon mass formula \cite{ioffe81}. 
We set $A=0$, and taking $M=m_n\sim $ 1~GeV, we divide the sum rule (\ref{sumrule})
by the standard $CP$-even sum rule for $\lambda$ obtained for the Lorentz structure $\psl$ 
and $\beta=1$ (the choice $\beta=-1$ in the latter sum rule leads to a similar result). 
The resulting estimate takes the following form, 
\ba
\label{ioffe_est}
d_n^{\rm est} 
 &=& \fr{8\pi^2|\qq|}{m_n^3}\left[ -\fr{2\chi m_*}{3}e(\bar\theta - \th_{\rm ind}) 
\right. \nonumber\\&& \;\;\;\;\;\;\;\;\;\;\;\;\;\;\;\;\;\;\;\;\;\left.
+\fr{1}{3}(4 d_d - d_u) 
+\fr{\chi m_0^2}{6}(4e_d\tilde d_d-e_u\tilde d_u)  \right],
\ea
where $\th_{\rm ind}$ again is a linear combination of $\tilde d_q/m_q$ (\ref{theta_ind}).
The coefficient in front of the square brackets in (\ref{ioffe_est}) is very close to 1, 
given Ioffe's estimate for $m_n$, $m_n^3 \simeq 8\pi^2|\qq|$ \cite{ioffe81}.
Indeed, this estimate shows no deviation at all from the naive quark model result for 
$d_n(d_q)$! Using Vainshtein's value for $\chi$, $\ch=-N_c/(4\pi^2f_\pi^2) \sim\ -9$ GeV$^{-2}$ 
\cite{vainshtein}, obtained
using pion-dominance for the longitudinal part of certain anomalous 
triangle diagrams, along with the Ioffe formula for $m_n$, 
the estimate for $d_n(\bar \theta)$ becomes
\ba
d_n^{\rm est} =  \fr{em_*\bar\theta}{2 \pi^2 f_\pi^2},
\ea 
which coincides with the chiral estimate (\ref{clog}) if $g_A \langle p| \bar q \tau^3 q |p\rangle 
\ln(\Lambda/m_\pi)$ is of order 2, where $g_A \simeq g_{\pi NN}f_\pi/m_n$. Needless to say that within
the accuracy of both methods the two estimates coincide. 
If $\bar\theta$ is removed by PQ symmetry, then within the same approximation 
the resulting estimate reads 
\ba
 d_n^{\rm est} = \fr 43d_d- \fr 13 d_u -
\frac{2m_\pi^2e}{m_n(m_u+m_d)}\left(\fr{2}{3}\tilde d_d + \fr 13 \tilde d_u\right),
\ea
where the approximate relation $m_0^2 \simeq -m_n^2$ has been used, along with 
$(m_u+m_d)|\qq|= f_\pi^2m_\pi^2$, are used. 
One immediately sees that the CEDM contributions 
are significant and comparable in magnitude in 
fact to the effects induced by quark 
EDMs.

We can give a more precise numerical treatment by making use of the following
parameter values: For the quark condensate, we take a central value of 
\ba
 \langle 0|\ov{q}q|0\rangle & = & - (0.225\mbox{ GeV})^3,
\ea
while for the condensate susceptibilities, we have \cite{chival}
\ba 
\label{chival}
 &&\ch \simeq  - 5.7 \pm 0.6 \mbox{ GeV}^{-2}, \nonumber\\ 
 && m_0^2  \simeq  -0.8 \mbox{ GeV}^2. 
\ea
This determination of $\ch$ is based on spectral sum-rules \cite{chival} and is slightly lower than
the value obtained in \cite{vainshtein}.

A more systematic treatment of the sum-rule \cite{pr2,pr3,pr5} indicates that a 
stability domain exists for relatively low Borel mass scales, of 
$M\sim {\cal O}(0.8 {\rm GeV})$. 
Convergence of the OPE is apparently not in danger as this low scale arises via the two 
step procedure used, in which the OPE is naturally formulated around the
neutron scale, $M\sim$1 GeV, while the chiral techniques used to
extract the dependence of the condensates on the $CP$-violating sources
lower the effective scale, but also introduce additional combinatoric suppression factors as the
dimension of the condensates increases. In order to test the stability of the sum rule, and obtain an
estimate for the uncertainty due to the handling of excited states, one
can generalise the expression for (\ref{sumrule}), by including a more systematic
parametrisation of the continuum, and also by including 
1-loop anomalous dimensions for the currents and condensates
entering the sum-rule. However, as alluded to earlier, these refinements
have a rather minimal impact, moving the stability domain by no more than
10-15\%. This is relatively small compared to the primary sources of
error, namely the saturation hypothesis for the condensates, the need to extract
the single-pole term from the sum-rules and, perhaps most significantly, the dependence 
on $\beta$. 

Extracting a numerical central value from the sum-rule, employing 
numerical estimates for the condensates (\ref{chival}),
and estimating the precision through 
consideration of the sources of error listed above, we find the results first 
presented in \cite{pr2,pr5}, 
\ba
\label{numerical}
 &&d_n(\bar\theta)= (1\pm 0.5) \frac{|\qq|}{(225 {\rm MeV})^3}\bar \theta \times 2.5 ~ 10^{-16}e\, {\rm  cm},\\
 && d^{\rm PQ}_{n}(d_q,\tilde{d}_q) = (1\pm 0.5) \frac{|\qq|}{(225 {\rm MeV})^3}
 \left[1.1e(\tilde d_d + 0.5\tilde d_u)+1.4(d_d-0.25d_u)\right],
 \nonumber
\ea
where we intentionally split the formula into two parts, $d_n(\bar\theta)$ and 
$d_n(d_q,\tilde d_q)$ in the presesence of PQ symmetry. In the generic case, 
the two lines in (\ref{numerical}) must be added together and $\bar \theta$ substituted by
$\bar \theta - \theta_{\rm ind}$. 

The result (\ref{numerical}) offers several interesting consequences. 
Note that the overall factor of $\qq$ combines with the 
light quark masses from short-distance expressions 
for $d_i$ and $\tilde d_i$ to give a result 
$\sim f_\pi^2 m_\pi^2(1+O(m_u/m_d))$ thus reducing the uncertainty 
due to poor knowledge of the absolute values of quark masses and condensates.
The constribution from the strange quark CEDM is suppressed by the $m_*/m_s$ ratio, 
and is completely removed at this order once the PQ symmetry is switched on. 
However, the use of nucleon currents with only valence quarks
leads one to suspect that gluon and sea-quark contributions may be under-estimated, since they
enter only at higher orders. It is possible that the question of $d_n(\tilde{d}_s)$ 
may be resolved in future lattice simulations, given an appropriate lattice implementation of
chiral symmetry.

Compared to the techniques outlined previously, this approach has the significant advantage that
all of the sources up to dimension five can be handled systematically and thus relative signs
and magnitudes can be consistently tracked. As indicated in (\ref{numerical}), one can also make a
systematic estimate of the precision of the result, where the errors are due to the contribution of excited states,
neglected higher dimensional operators in the OPE, and also an ambiguity in the nucleon current.
Comparing the numerical result with those obtained using NDA and chiral techniques one finds, 
as is to be expected, that the results agree in terms of order of magnitude. 
In particular, our result for $d_n$ implies a bound 
$|\bar{\theta}|< 3\times 10^{-10}$.

In progressing to consider the contribution from sources of 
higher dimension, problems arise through the appearance of certain infrared divergences 
at low orders in the OPE, while a number of unknown condensates also enter and render 
a corresponding calculation for dimension six sources intractable.
One can nonetheless estimate the contribution of these operators by utilizing a trick 
which involves relating the EDM contribution to the measured anomalous magnetic
moment $\mu_n$ via the $\ga_5$--rotation of the nucleon wavefunction induced by
the $CP$-odd source \cite{BU},
\begin{eqnarray} 
   d_n\sim \mu_n\,\fr{\langle N\vert \de {\cal L}_{CP}\vert N\rangle} {m_n\bar N i\gamma_5 N}.
   \label{BU}
\end{eqnarray}
One may analyze the $\ga_5$-rotation using conventional sum-rules techniques, and for the
Weinberg operator, one can obtain the following estimate \cite{DPR}
\ba
  |d_n(w)| \sim |\mu_n|  \fr{3 g_s m_0^2}{32 \pi^2} w\ln{(M^2/\mu_{IR}^2)} 
\simeq e~ 22~{\rm MeV}~ w(1~{\rm GeV}),
\label{est2}
\ea 
taking $M/\mu_{IR} = 2$, where $M$ is the Borel mass and $\mu_{IR}$ is an infrared cutoff,
and $g_s = 2.1$.

We can also apply this technique for the contribution of four-fermion operators.
For SUSY models with generic parameters $CP$-odd four-fermion operators
are negligible due to the double helicity-flip requiring an $m_q^2$ dependence
and rendering these operators effectively of dimension eight. However,
for large $\tan\beta$, there are enhancements for operators proportional to $C_{ij}$ 
with $i,j=d,s,b$ which can partially 
overcome this suppression thus altering the conventional picture of EDM sources
(see Fig.~\ref{schemetb}). An important class of contributions in this case 
involves the four-ferimon operators with a $b$-quark. The contribution of 
these sources to $d_n$ can again be estimated using the same technique as above \cite{dlopr},
\ba
 |d_n(C_{ij})| \sim e\ 2.6 \times 10^{-3} {\rm GeV}^2 
\left(\frac{C_{bd}(m_b)}{m_b} 
 + 0.75 \frac{C_{db}(m_b)}{m_b}\right).
\ea

We should emphasize that both the dimension six estimates above necessarily have a
precision not better than ${\cal O}(100\%)$, and one cannot reliably extract
the sign. Fortunately, the numerical size of these dimension six contributions is often
negligible, and thus does not significantly impact the phenomenological application of
EDM constraints.

\bigskip
$\bullet$ {\it Calculation of $\bar g_{\pi NN}$} 
\bigskip

The other primary source of nuclear EDMs, leading to the observable EDMs in
diamagnetic atoms, arises through nucleon interactions mediated by pion-exchange with 
$CP$ violation in the pion-nucleon vertex. As discussed in the preceding subsection, the calculation
of these couplings involves essentially two steps: the first is a PCAC-type reduction of the
pion in $\langle N \pi^a | {\cal O}_{CP} |N^\prime\rangle$ as in (\ref{start}), and 
the second is an evaluation of the resulting matrix elements over the nucleon. It is this second
part for which QCD sum-rules may usefully be employed, and here we sketch its application to
the computation of the dependence of $\bar{g}_{\pi NN}$ on dimension five $CP$-odd sources
in (\ref{manyedms}) \cite{pospelov01}.

The main difficulty, as alluded to in the previous section, is the partial cancellation between 
the $m_0^2 \bar qq$ and $ \bar q g_s G\si q$ sources in (\ref{chiral_g}). Within the sum-rule analysis, one 
can nonetheless trace this cancellation up to the next-to-next-to-leading order \cite{pospelov01}.
Indeed, following the approach outlined earlier of fixing the unphysical parameter $\beta$ to suppress the
highest calculated order in the OPE, leads again to $\beta=1$. However, in this case, the choice $\beta=1$
sets all the LO, NLO, and NNLO terms to zero! This cancellation is, however, seemingly an artefact of the 
purely valence-quark content of the current, and is not dictated by symmetry. Thus, although the result
will necessarily have a strong dependence on $\beta$. one can obtain a numerical estimate by varying
$\beta$ through an appropriate range. To get some idea of the size of the result, let us work only
with the current $\et_1$ which survives in the nonrelativistic limit and has the dominant overlap
with the neutron state, i.e. set $\beta=0$. Assuming also  the dominance of the nucleon double-pole term
and the leading order term in the OPE, and separating different isospin structures, we find \cite{pospelov01}
\ba
\gone \sim \fr{3}{4}~
\fr {m_0^2}{f_\pi}(\tilde d_u - \tilde d_d),~~~~
\gzero \sim \fr{3}{20}~
\fr {m_0^2}{f_\pi}(\tilde d_u +\tilde d_d).
\ea
Unless $\tilde d_u - \tilde d_d \simeq 0$, the $CP$-odd 
coupling $\gone$ is several times larger than $\gzero$.
Concentrating therefore on $\gone$, we have
\ba
\label{nnum1}
\gone = 3\times10^{-12}\fr{\tilde d_u- \tilde d_d}{10^{-26} {\rm cm}}
\fr{|\qq|}{(225 {\rm MeV})^3}~\fr{|m_0^2|}{0.8{\rm GeV}^2},
\ea
where we re-instate the dependence on the relevant condensates,
normalized to their central values. The estimate (\ref{nnum1}) is 
half the size of the value for $\gone$ used in \cite{FOPR}.  

The full numerical treatment of the sum-rule for 
the isospin-one coupling, produces the following
result at next-to-next-to-leading order \cite{pospelov01},
\ba
\label{nnum2}
\gone = 2^{+4}_{-1} \times10^{-12}\fr{\tilde d_u- \tilde d_d}{10^{-26} {\rm cm}}
\fr{|\qq|}{(225 {\rm MeV})^3}, \label{goneres}
\ea
where the poor precision is essentially due to the cancellation of the leading terms as described
above. We emphasize that a more precise calculation of the matrix element (\ref{thisiswhatweneed}) would significantly 
enhance the quality of constraints one could draw from the experimental bounds on diamagnetic EDMs, and thus
constitutes a significant outstanding problem. Again, it seems progress may  have to wait
for further developments in lattice QCD.

In considering the contribution of dimension six sources to $\bar{g}_{\pi NN}$, we note
firstly that the three-gluon Weinberg operator $GG\tilde G$
is additionally suppressed by $m_q$ and can be neglected. 
However, as for $d_n$, for SUSY models with large $\tan\beta$, certain four-fermion operators
$C_{q_1q_2}$ may be relevant, and can be obtained via vacuum factorization, as the two diagrams
in Fig.~3 now fail to cancel, 
\ba
\bar g_{\pi NN}^{(1)}
 \sim -8\times 10^{-2}{\rm GeV}^2\left(\fr{0.5C_{dd}}{m_d} 
+ 3.3\kappa\fr{C_{sd}}{m_s}+ 
\fr{C_{bd}}{m_b}(1-0.25\kappa)\right),
\label{gone}
\ea
where $\ka\equiv\langle N| m_s \bar{s}s |N\rangle/\mbox{ 220 MeV}$, with the preferred value $\kappa \simeq 0.5$
\cite{KBM}.

\section{EDMs in models of $CP$ violation}

We have now moved to the highest level in Fig.~1, 
which is where the EDM constraints can 
be applied to directly constrain new sources of $CP$ violation. In this section, we will breifly discuss
these constraints, firstly looking at why the Standard Model itself provides such a small background, and then
why most models of new physics, and supersymmetry in particular, tend to overproduce EDMs and are thus subject
to stringent constraints.

\subsection{EDMs in the Standard Model}

The recent discovery and exploration of $CP$ violation in the neutral $B$-meson system
\cite{sin2beta} is, along with existing data from $CP$-violation observed in $K$-mesons, 
(within current precision) in accord with the minimal model of $CP$ violation known as the 
Kobayashi-Maskawa (KM) mechanism \cite{KM}. This introduces a $3\times3$ unitary quark mixing 
matrix $V$ in the charged current sector of up and down-type 
quarks taken in the mass eigenstate basis,
\ba
{\cal L}_{cc} = \frac{g}{\sqrt{2}}\left( \bar U_L \Wsla^{\,\,+} VD_L \, + \,
\mbox{(H.c.)}\right). \label{cc}
\ea
This model possesses a single $CP$-violating invariant in the quark sector, $J_{CP}=$
Im$(V_{tb}V^*_{td}V_{cd}V^*_{cb}) \simeq 3 \times 10^{-5}$. 
This combination, along with $\theta_{\rm QCD}$, 
are the only allowed sources of $CP$ violation in the Standard Model (treating ``Standard Model neutrinos'' 
as massless). In addition to this, $CP$ violation in the SM vanishes in the limit of an equal 
mass for any pair of quarks of the same isospin, e.g. $d$ and $s$, $u$ and $c$, etc.
These two conditions are extremely powerful in suppressing any KM-induced
$CP$-odd flavour-conserving amplitude. 

\begin{figure}[t]
 \centerline{%
   \includegraphics[width=6cm]{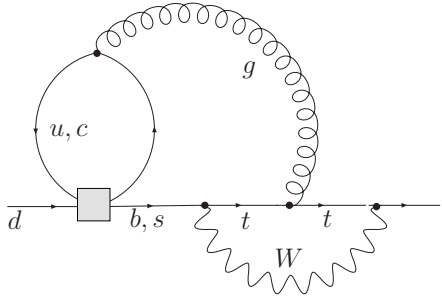}%
         }
 \caption{A particular 3-loop contribution \protect\cite{CK} 
to the $d$-quark EDM induced by the KM phase in the standard model.
The box vertex denotes a contacted $W$-boson line connected to the light quarks, while it is implicit that 
the external photon line is to be attached as appropriate to any charged lines.}
 \label{SModel}
\end{figure}

\bigskip
$\bullet$ {\it quark and nucleon EDMs}
\bigskip

The necessity of four electroweak vertices requires
that any diagram capable of inducing a quark EDM have at
least two loops. Moreover, it turns out that all EDMs and color
EDMs of quarks vanish exactly at the two-loop level \cite{Shab},
and only three-loop diagrams survive \cite{Kh,CK}, as in Fig.~\ref{SModel}.
A leading-log calculation of the three-loop amplitude for the EDM of the
$d$-quark produces the following result \cite{CK}, 
\ba
d_d = e \fr{m_dm_c^2\alpha_s G_F^2 J_{CP}}{108\pi^5}
\ln^2(m_b^2/m_c^2)\ln(M_W^2/m_b^2).
\ea
Upon the inclusion of the other contributions, it produces a 
numerical estimate
\ba
d_d^{\rm KM} \simeq 10^{-34}e~{\rm cm}.
\ea

\begin{figure}[t]
 \centerline{%
   \includegraphics[width=12cm]{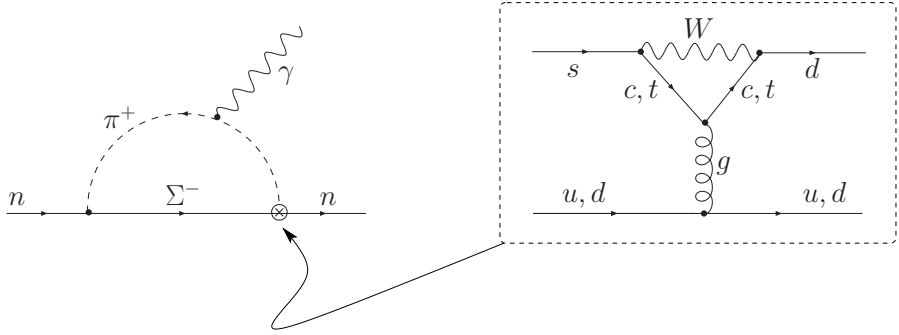}%
         }
 \caption{A leading contribution to the neutron EDM in the Standard Model, arising via a four-quark operator
generated by a strong penguin, and then a subsequent enhancement via a chiral $\pi^+$ loop.}
\label{strongP}
\end{figure}

The only relevant operator that is not zero at two-loop order
is the Weinberg operator \cite{W-P}, but its numerical value also turns
out to be extremely small. Indeed the largest Standard Model contributon to $d_n$ 
comes not from quark EDMs and CEDMs, but instead from a four-quark operator generated by a so-called 
``strong penguin'' diagram shown in Fig.~\ref{strongP}. This is enhanced by long distance effects, namely the
pion loop, and it has been estimated that this mechanism could lead to  a KM-generated EDM
of the neutron of order \cite{LD},
\ba
 d_n^{\rm KM} \simeq 10^{-32}e~{\rm cm}. 
\ea
However, this is still six to seven orders of magnitude smaller than the current
experimental limit. 

\bigskip
$\bullet$ {\it lepton EDMs}
\bigskip

The KM phase in the quark sector can induce a lepton EDM
via a diagram with a closed quark loop, but a
non-vanishing result appears first at the four-loop level \cite{KP}
and therefore is even more suppressed, below the level of 
\ba
 d_e^{\rm KM} \leq 10^{-38}e~{\rm cm},
\ea
and so small that the EDMs of paramagnetic atoms and molecules would 
be induced more efficiently by e.g. Schiff moments and other $CP$-odd nuclear momenta.  

In this regard, we note that recent data on neutrino oscillations points toward the existence 
of neutrino masses, mixing angles, and possibly of new CKM-like phase(s) in the lepton 
sector. Under the assumption that neutrinos are Majorana particles, the presence of these new $CP$-odd 
phases in the lepton sector allows for a non-vanishing two-loop contributions to $d_e$ \cite{ngng},
without any further additions to the Standard Model. However, recent calculations \cite{ACP} 
show that a typical see-saw pattern for neutrino masses and mixings only induces a tiny contribution 
to the EDMs in this way, of $O(m_em_\nu^2G_F^2)$, unless a fine-tuning of the light neutrino masses is tolerated 
in which case $d_e$ could reach $10^{-33}e~$cm. Therefore, within this minimal extension of the Standard Model
allowing for massive neutrinos, the electron EDM is not the best way to probe $CP$ violation in 
the lepton sector. 

\bigskip
$\bullet$ {\it Probing the scale of new physics}
\bigskip

The Standard Model predictions for EDMs described above are well beyond the reach of 
even the most daring experimental proposals. This implies in turn that the Kobayashi-Maskawa phase 
provides a negligible background and thus any positive detection of an EDM would necessarily imply 
the presence of a non-KM $CP$-violating source. Before we consider some of the models which provide
motivations for anticipating such a discovery, it will first be useful to consider in more general terms
how high an energy scale one could indirectly probe with EDM meaurements. Indeed, we are led to
ask first of all, what energy scale of new $CP$-violating physics is probed with the current
experimental sensitivity to EDMs? Secondly, given the small KM background, we might also ask for the
largest energy scale that could be probed {\it in principle} before reaching the level where the
Standard Model KM contibutions would become significant.

To try and answer these questions in a systematic way, let us consider a toy model containing a  
scalar field $\phi$ (which is Higgs-like, but needn't be the SM Higgs) coupled to the SM fermions
with scalar and pseudoscalar couplings, 
${\cal L}_{\rm int} = - \sum_{i}\phi\bar \psi_i (y_i +iz_i\gamma_5)\psi_i$. 
The presence of both 
scalar and pseudoscalar couplings $y_i$ and $z_i$ breaks 
$CP$ invariance.

Assuming that the scalar mass $M$ is large, we integrate this field out and match the 
resulting coefficients with the Wilson coefficients listed in (\ref{manyedms}) and (\ref{manydim6}).
In particular, at tree level, 
we obtain the following contribution to (\ref{eN}),
\ba
C_S^{(0)} \simeq \fr{1}{M^2}z_e( 3(y_d + y_u) + \kappa y_s+\cdots), \label{cij}
\ea
where the ellipsis stands for the contribution of heavy quark flavours,
and the couplings are normalized at 1 GeV. If we now make the assumption that there is no correlation with 
other sources of $CP$ violation, e.g. the electron EDM $d_e$, then with the use of the experimental constraint 
on $d_{\rm Tl}$ and the results of atomic claculations (\ref{dtl}), we arrive at the 
following limit on the $CP$-odd combination of couplings and the mass $M$, 
\ba
\fr{1}{M^2}z_e( y_d + y_u + 0.3\kappa y_s)\leq \fr{1}{(1.5\times 10^6 {\rm GeV})^2}.
\ea
Given the most optimistic assumption about the possible size of these couplings, i.e. $z_ey_q \sim O(1)$, 
we can conclude that the current experimental EDM sensitivity translates to a bound on $M$ as 
high as $M_{\rm CP} \sim 10^6$ GeV. If instead we insert the largest Kobayashi-Maskawa prediction for 
the Tl EDM of order $\sim 10^{-35}e$ cm in place of the current sensitivity, we 
obtain $M_{\rm CP}^{\rm max} \sim 10^{11}$ GeV, as the ulmitate scale which can be probed via 
these dimension 6 operators before the onset of the ``KM background''. 

For comparison, allowing for arbitrarily large flavour-violating couplings of $\phi$ to fermions, 
we can also deduce the sensitivity level to New Flavour Physics (NFP) in a similar way. 
For example, requiring that four-fermion operators that change flavour by two units, 
e.g. $\bar s \gamma_5 d \bar s \gamma_5d$ and the like, do not introduce new contributions to 
$\Delta m_B$, $\Delta m_K$ and $\epsilon_K$ that are larger than the SM contribution, one typically 
finds that $M_{\rm NFP} > 10^7-10^8$ GeV in this scenario. Thus, we see that the sensitivity of EDMs is
already approaching this benchmark and, unlike the contraints from the $\Delta F=2$ sector, 
can be significantly improved in the near future.

At this point, we should emphasize that in this example, we are relaxing all constraints
on the flavour structure by allowing order one couplings of the scalar field $\phi$ to the light fermions.
These couplings violate chirality maximally, and if 
they were part of a more realistic 
construction, for example a two-Higgs doublet model, one would expect that $y_i$ and $z_i$ will have to 
scale according to the fermion mass $m_i$. In this case, the sensitivity to $M$ clearly 
drops dramatically, and the tree-level interactions (\ref{cij}) are not necessarily the dominant 
contributions to EDMs, as heavy flavours may contribute in a more substantial way via loop effects \cite{W,BZ}. 
Indeed, if new physics above the electroweak scale preserves chirality, as is often assumed,
one expects that for the light flavours  $d_i \sim e\times (1-10)\,{\rm MeV}/M^2$. Taking the
electron EDM, and the Tl EDM bound, as a concrete example we find under this more restrictive
assumption that
\ba
 d_e \sim e\times \frac{m_e}{M^2}= 10^{-23}e\,{\rm cm}\times\left(\fr{1\,{\rm TeV}}{M}\right)^2 \;\;\;\;
    \Longrightarrow \;\;\;\; M_{\rm CP} \sim 70\, {\rm TeV},
    \label{mcp}
\ea
and consequently the current level of sensitivity to new $CP$-violating chirality-preserving physics drops 
somewhat, but for reference this scale is still well-beyond the centre-of-mass
energy of the LHC.

If we put the current EDM bounds into the broader context of precision tests of the Standard Model, 
we see that the present bounds in Table~1 imply that EDMs occupy an intermediate position in 
sensitivity to mass scales for new physics, between the electroweak precision tests (EWPTs) and 
flavour violation in $\Delta F =2$ processes noted above. The EWPTs from LEP impose 
a bound, $M_{\rm EWPT} > $ few TeV, through constraints on various dimension six operators, 
e.g. oblique corrections to gauge boson propagators. Since $M_{\rm EWPT} \gg M_Z$, this has been dubbed the 
``little hierarchy problem'' \cite{barbieri}. Indeed, while there are general
expectations that the Standard Model is an effective theory, and will be corrected  
at scales of about a TeV, it is clear from this discussion that precision constraints in 
many sectors do not contain any hints of new physics beyond the Higgs at the weak scale, and in this 
sense EDMs are no exception. The remarkably large scale $M_{\rm CP}$ implied by EDM limits 
requires, at least within our current level of understanding, a tuning in the $CP$-odd sector of 
physics beyond the standard model that we lack a coherent explanation for. 
The recent data from BaBar and Belle on $CP$ violation in the neutral $B$-meson sector, which thus
far is consistent with the KM model, within which $CP$ violation is maximal within the confines of
the flavour structure, only makes this tuning more pronounced, since we lack a strong motivation
to enforce any additional $CP$-violating phases to be small. Moreover, further
experimental progress in the near future could, given null results, push the value of 
$M_{\rm CP}$ closer to $M_{\rm NFP}$. From this viewpoint EDMs provide our most powerful tool 
in probing the question of whether $CP$-violation and flavour physics are 
intrinsically linked, as indeed they are within the electroweak Standard Model. This issue stands out as one
of the most important ways in which EDMs may assist in demystifying some of the less constrained
parts of the Standard Model.

\subsection{EDMs in supersymmetric models}

Having demonstrated the generic importance of EDM constraints for TeV-scale physics, we would now
like to make this analysis more concrete by focussing on models with electroweak scale 
superymmetry and reviewing their predictions for EDMs (see e.g. \cite{modrevs} for reviews of EDMs 
within several other classes of models).

Supersymmetric extensions of the Standard Model provide perhaps the most natural 
solution to the gauge hierearchy problem by automatically cancelling the quadratically 
divergent contributions to the Higgs mass. Supersymmetry is thought of here as a symmetry of nature 
at high energies, whereas at the electroweak scale and below it is obviously broken. 
Ensuring that supersymmetry breaking does not re-introduce quadratic divergences, and is 
compatible with the observed low energy spectrum, still allows for a large number of new 
dimensionful parameters, unfixed by any symmetry, that are usually called 
the soft breaking parameters. The minimal realization, known as the Minimal Supersymmetric
Standard Model (MSSM), has been the subject of numerous theoretical studies, and also experimental 
searches, for over two decades. While no experimental evidence for SUSY exists, the MSSM retains a
pre-eminent status among models of TeV-scale physics in part through several indirect virtues, e.g. gauge 
coupling unification and a ``natural'' dark matter candidate. For full details of the MSSM spectrum 
and the parametrization of the soft-breaking terms, we address the reader to any of the comprehensive 
reviews on MSSM phenomenology \cite{MSSM}. 

The unbroken sector of the MSSM contains, besides the gauge interactions, 
the Yukawa couplings parametrized by 3$\times$3 Yukawa matrices in flavour space, 
${\bf Y}_u$, ${\bf Y}_d$ and ${\bf Y}_e$. These matrices source the
tree-level masses of matter fermions, 
\ba 
{\bf M}_u = {\bf Y}_u \langle H_2\rangle,\;
{\bf M}_d = {\bf Y}_d \langle H_1\rangle,\;
{\bf M}_e = {\bf Y}_e \langle H_1\rangle,
\ea
where $\langle H_1\rangle$ and $\langle H_2\rangle$ are two Higgs vacuum expectation values
related to the SM Higgs v.e.v. via  $ \langle H_2\rangle^2+\langle H_1\rangle^2 = v^2/2$. 
In SUSY models, anomaly cancellation in the Higgsino sector requires the introduction of 
at least two Higgs superfields as above. In addition to Yukawa couplings, the supersymmetry-preserving 
sector contains the so-called $\mu$-term that provides a Dirac mass to the higgsinos (the 
superpartners of the Higgs bosons) and contributes to the mass term of the Higgs potential,
\ba
V_{\rm Higgs} = m^2_1|H_1|^2 +  m^2_2|H_2|^2 +m^2_{12}H_1H_2+ |\mu|^2(|H_1|^2+|H_2|^2)+\cdots, 
\ea 
where the ellipses denotes quartic terms fixed by supersymmetry and gauge invariance 
\cite{MSSM}. $m^2_1$, $m^2_2$ and $m^2_{12}$ are soft-breaking parameters that may attain 
negative values thus driving electroweak symmetry breaking. By suitable phase redefinitions 
of $H_1$ and $H_2$, one can restrict to real Higgs v.e.v.s and introduce the parameter,
$\tan\beta=\langle H_2\rangle/\langle H_1\rangle $.

Among the remaining soft-breaking parameters one has  
gaugino mass terms and squark and slepton masses,
\ba 
-{\cal L}_{\rm mass} = \fr{1}{2}\sum_{i=1,2,3}M_i\bar\lambda_i\lambda_i 
 + \sum_{S=Q,U,D,L,E}\tilde{S}^{\dagger}{\bf M}^2_{\tilde S}\tilde S,
\ea
where $\lambda_i$ are the gaugino (Majorana) spinors, with $i$ labelling the 
corresponding gauge group, U(1), SU(2) or SU(3). Each gaugino mass $M_i$ can be complex.  
The second sum spans all the squarks and sleptons and contains five Hermitian 
3$\times$3 mass matrices in flavour space. 
Finally, the soft-breaking terms also include three-boson couplings
allowed by gauge invariance, such as $\tilde{Q}H_2{\bf A}_u\tilde{U}$,
that are called $A$-terms and are parametrized by three arbitrary 
complex matrices ${\bf A}_u$, ${\bf A}_d$ and ${\bf A}_e$.
In the construction above we have limited the discussion to 
the $R$-parity conserving case, which only allows an {\em even} number of 
superpartners in each physical vertex, and is imposed to reduce problems with baryon number 
violation. Even with this restriction, if we count all the free parameters in this model we find
a huge number, of ${\cal O}(100)$, with a few dozen new $CP$-violating phases! Truncating this
number is fully justified only within the context of a fully specified supersymmetry breaking
mechanism, which may then enforce additional symmetries and relations among parameters.

Without going into the details of the dynamics behind SUSY breaking, it will be enough for
our purposes to simply assume that the following, very restrictive, conditions are fulfilled: 
\begin{eqnarray}  
\nonumber
{\bf M}_S^2= m_S^2{\bf 1};\;\;{\rm for} ~ S =Q,U,D,L,E,
\;\;\;\mbox{``degeneracy''} \\
{\bf A}_i= A_i{\bf Y}_i;\;\;{\rm for} ~ i =u,d,e,
\;\;\;
\mbox{``proportionality''},  \label{prop}
\end{eqnarray}
Strictly speaking, such conditions can only be imposed 
at a specific normalization point above the weak scale, as the renormalization 
group evolution of the MSSM parameters will modify these relations. Moreover, these conditions
can only be imposed with limited precision at this scale due to threshold effects. Nonetheless, 
such a restrictive flavour universality ansatz in the scalar mass sector, and 
proportionality of the trilinear soft breaking terms to the Yukawa matrices has the utility that
it greatly reduces the number of independent soft-breaking parameters. Even so, 
a significant number of $CP$-violating phases remain, e.g.
\ba
{\rm Arg}(\mu M_i m_{12}^{2*}),~ {\rm Arg}(A_f M_i^* ),~
{\rm Arg}(M_i M_j^*),~ {\rm Arg}(A_{f} A_{f'}^*).
\ea
Going to an even more restrictive framework, by assuming a common phase for 
the gaugino masses and another common phase for $A_i$ reduces the 
number of independent $CP$ violating parameters to two. Using phase redefinitions, 
one can choose the phase of the gaugino mass to be zero, and use 
$\theta_A = {\rm Arg}(A)$ and $\theta_\mu = {\rm Arg}(\mu)$
as the basis for parametrizing $CP$ violation.

It has been known for over twenty years that even in the absence of
new flavour physics, large EDMs can be induced at the one-loop level within a single genaration
\cite{WWI,Hall}. Indeed, one would anticipate large EDMs as both of the reasons that rendered
$d_i(\delta_{KM}$) very small, namely high loop order and also mixing angle/Yukawa 
coupling suppression, are not present for EDMs induced by the phases of the soft-breaking 
parameters.

\begin{figure}[t]
 \centerline{%
   \includegraphics[width=12cm]{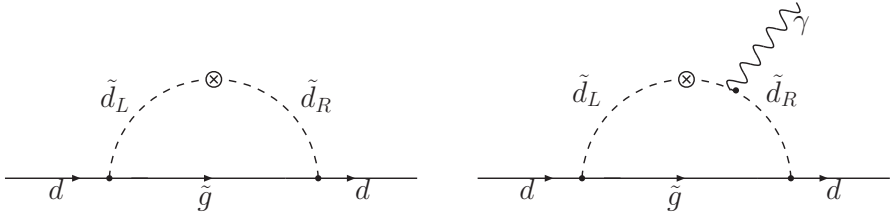}
         }
\vspace{0.1in}
 \caption{\small One-loop SUSY threshold corrections in the down quark sector induced
by a gluino-squark loop. On the left, a threshold correction generating Im$(m_d)$, while on
the right the analogous diagram for the EDM. The $CP$-violating source enters via the highlighted 
vertex, squark-mixing in the present case. }
\end{figure}

Figure~7 exhibits  examples of one-loop  diagrams  at the supersymmetric threshold that 
generate non-zero contributions to the $CP$-odd Lagrangians (\ref{thetaterm}) and (\ref{manyedms}). 
If we leave aside the problematic $s$-quark CEDM, then at one-loop we can concentrate on  
diagrams involving just the first generation of quarks and leptons. 
Within the parametrization described above, the phases residing in $\mu$ and $A$ permeate the squark,
selectron, chargino and neutralino spectrum, which in the mass eigenstate basis 
translates into complex phases in the quark--squark--gluino and 
fermion--sfermion-chargino(neutralino) vertices. To make this explicit, for a moment let us 
truncate the flavour space to one generation and write down the expression for the 2$\times$2 
$d$-squark mass matrix at the electroweak scale in the basis of $\tilde d_L$ and $\tilde d_R$,
\ba
M_{\tilde{d}}^2 =  \left(\begin{array}{cc}
                                     m_Q^2 +O(v^2) & -m_d (\mu \tan\beta + A_d^* )\\
                                      -m_d (\mu^* \tan\beta + A_d) & m_D^2 + O(v^2)\end{array}\right),
                                      \label{2by2}
\ea
where we further assume that the soft masses $m_{Q}^2$ and $m_{D}^2 $
are large relative to the weak scale, and thus we can ignore subleading $O(v^2)$ corrections 
to the diagonal entries. Similar expressions can be written for the selectron mass matrix 
with the obvious substitutions in (\ref{2by2}), and for the $u$ squark, where in addition 
one has to exchange $\tan \beta$ by $\cot \beta$. In the generic case of three generations, 
$M^2$ becomes a 6$\times$6 matrix with 3$\times$3 blocks which are traditionally 
called $M^2_{LL}$, $M^2_{LR}$, $M^2_{RL}$ and $M^2_{RR}$. For our purposes, the crucial terms in (\ref{2by2}) 
are the off-diagonal components, 
\ba
 (M_{\tilde d}^2)_{LR} = -m_d (\mu \tan\beta + A^*_d)
\ea
which contain the $CP$-odd phases. By virtue of being proportional to the small mass $m_d$, 
such a term can be treated as a perturbation and accounted for by an explicit mass insertion 
on the squark line, as in Figure 7. Note that the natural range for $\tan\beta$, in the 
interval between 1 and 60, allows for a significant enhancement of the $\mu$-dependent 
term in (\ref{2by2}). The phase of $\mu$ also modifies the spectrum of charginos and neutralinos,
the mass eigenstates of the superpartners of the $Z$, $W$, $\gamma$ and Higgs bosons. 

With this notation in hand we see that, for example, the squark--gluino loop diagram 
generates an imaginary $d$-quark mass correction that contributes to $\Delta \bar \theta_{\rm rad}$,
\ba
{\rm Im} \; m_d = -m_d \fr{\alpha_s}{3\pi}
\fr{M_3(\mu\tan\beta \sin\th_\mu-A_d\sin\th_A)}{M_Q^2}I(M_3,m_Q,m_D).
 \label{oneloopth}
\ea
The loop function $I$ is normalized in such a way that 
$I(m,m,m)=1$; its exact form (see e.g. \cite{dlopr,IN}) is not important for our discussion. 
The ratio ${\rm Im} \, (m_d)/m_d$, along with contributions from other quark flavours, represent 
a one-loop renormalization of the $\th$--term. It is important to observe that it only depends
on the SUSY mass ratio and thus does not decouple if $A, \mu, M_3, m_{Q(D)}$ are pushed far above the 
electroweak scale. Applying the bound on the $\th$--term to the combined tree level and 
one-loop results (\ref{oneloopth}), with degenerate SUSY mass parameters as above,
we find $|\bar\theta_{\rm tree} + 10^{-2}\de_{\rm CP}|< 10^{-9}$, where $\delta_{\rm CP}$ is a linear 
combination of $\sin\theta_\mu$ and $\sin\theta_A$ with ${\cal O}(1)$ coefficients.
If there is no axion and $\bar\theta_{\rm tree}$ vanishes instead by symmetry arguments, 
it follows that the phases of the soft-breaking parameters must be tuned to within a 
factor of $10^{-7}$ in order to satisfy the EDM bounds. Therefore, an incredibly tight 
constraint on the phases of the SUSY soft-breaking parameters can be obtained in 
models which invoke high-scale symmetries to resolve the strong $CP$ problem.

However, if the PQ symmetry removes the $\th$--term, such radiative corrections to $\bar\theta$
have no physical consequences, and the residual EDMs are determined by higher dimensional operators.
The relevant expressions for the one-loop--induced $d_i$ and $\tilde d_i$ contributions 
can be found in {\em e.g.} \cite{IN}. Here we would just like to demonstrate the main point implied
by these SUSY EDM calculations in a simplistic model in which {\em all} soft-breaking parameters
are taken equal to a unique scale $M_{\rm SUSY}$ at the electroweak scale, i.e. 
$M_i=m_Q=m_D=\cdots=|\mu|=|A_i|=M_{\rm SUSY}$. Working at leading order in $v^2/M^2_{\rm SUSY}$, 
we can then present the following compact results for all dimension 5 operators (with $q=d,u$), 
\begin{eqnarray}
\nonumber
\fr{d_e}{e\kappa_e} &=& \frac{g_1^2}{12} \sin\theta_A+ \left(\frac{5g_2^2}{24}+
\frac{g_1^2}{24}\right)\sin\theta_\mu\tan\beta   ,\\ 
\label{simple_d}
\fr{d_q}{e_q\kappa_q}&=& \frac{2g_3^2}{9}\Big(\sin\theta_\mu[\tan\beta]^{\pm1}-\sin\theta_A\Big) + O(g_2^2,g_1^2),\\
\fr{\tilde d_q}{\kappa_q} &=&\frac{5g_3^2}{18} \Big(\sin\theta_\mu[\tan\beta]^{\pm1}-\sin\theta_A\Big)+ O(g_2^2,g_1^2).
\nonumber
\end{eqnarray}
The notation $[\tan\beta]^{\pm1}$ implies that one uses the plus(minus) sign for $d$($u$) quarks, 
$g_i$ are the gauge couplings, and $e_u=2e/3$, $e_d=-e/3$. For the quarks we quoted the explicit result only
for the gluino-squark diagram that dominates in this limit. All these contributions to $d_i$ are 
proportional to $\kappa_i$, a universal combination corresponding to the generic dipole size,
\ba
\kappa_i = \fr{m_i}{16 \pi^2 M^2_{\rm SUSY}} = 1.3\times 10^{-25} 
{\rm cm}\times \fr{m_i}{\rm 1MeV} \left(\fr{1{\rm TeV}}{M_{\rm SUSY}}\right)^2,
\label{universal}
\ea
which varies by a factor of a few for $i=e,d,u$ depending on the value of the fermion mass.  
The perturbative nature of the MSSM provides a loop suppression factor in (\ref{universal}) 
so that $\kappa_i$ is about two orders of magnitude smaller than the estimate (\ref{mcp}).
Correspondingly, the reach of the current EDM constraints in SUSY models cannot exceed the scale
of a few TeV. 

In (\ref{universal}) the quark masses should be normalized at the high scale, $M_{\rm SUSY}$.
To make the explicit connection with the dipole operators in (\ref{manyedms}), the results of 
Eq.~(\ref{simple_d}) should be evolved down to the low-energy normalization point of 1 GeV using 
the relevant anomalous dimensions (see e.g. \cite{dlopr}). Plugging these results into 
the expressions for $d_n$, $d_{\rm Tl}$ and $d_{\rm Hg}$ and comparing them to the 
current experimental bounds, we arrive at a set of constraints on $\theta_A$ and $\theta_\mu$ depending
on  $M_{\rm SUSY}$ and $\tan\beta$. In Figure~8, we plot these constraints in the 
($\theta_\mu$,$\theta_A$)--plane for $M_{\rm SUSY}=500$ GeV and 
$\tan\beta = 3$. The region allowed by the EDM constraints is at the intersection
of all three bands around $\theta_A=\theta_\mu=0$. One can observe that the combination of all 
three constraints strengthens the bounds on the phases, and protects against the accidental 
cancellation of large phases that can occur within one particular observable. 
The uncertainty in the QCD calculations of $\bar g^{(1)}_{\pi NN}$ and the nuclear calculation of 
$S(g^{(1)}_{\pi NN})$ discussed earlier may affect the width of the $d_{\rm Hg}$ constraint 
band, but do not change its slope on the ($\theta_\mu,\theta_A$) plane.

\begin{figure}[t]
 \centerline{%
   \includegraphics[width=5.5cm]{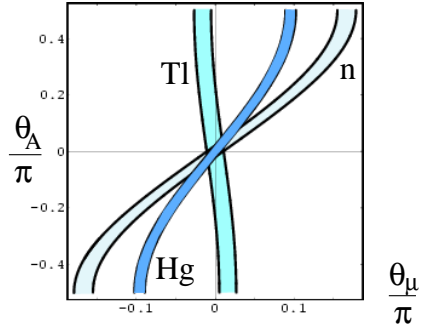}
         }
\vspace{0.1in}
 \caption{\small The combination of the three most sensitive EDM constraints, 
$d_n$, $d_{\rm Tl}$ and $d_{\rm Hg}$, 
for $M_{\rm SUSY} = 500$ GeV, and $\tan\beta=3$. The region allowed by EDM constraints is at the intersection
of all three bands around $\theta_A=\theta_\mu=0$. }
\end{figure}

Before we review the most common approaches to address the ``overproduction'' 
of EDMs in supersymmetric models, for completeness, we will briefly discuss some of the
additional contributions which become important when $\tan\beta$ is large, a regime
favoured for consistency of the MSSM Higgs sector with the final LEP results \cite{LEPref}.
One simple observation is that the EDMs of down quarks and electrons,  
induced by $\theta_\mu$ at one-loop, grow linearly  with $\tan\beta$ (\ref{simple_d}).
However, at the two-loop level, there are additional contributions from the phase of the $A$-parameter which
may also be $\tan\beta$--enhanced \cite{twoloop}. A typical representative of the two-loop family 
is presented in Figure~9. At large $\tan\beta$ the additional loop factor can be overcome, and 
these two-loop effects have to be taken into account alongside 
the one-loop contributions in (\ref{simple_d}). For example, the stop-loop contribution to the 
electron EDM in the same limit of a large universal SUSY mass is given by
\ba
d_e^{\rm two~loop} = - e\kappa_e \fr{\alpha Y_t^2}{9\pi}\ln\left[\fr{M_{\rm SUSY}^2}{m_A^2}\right]
\sin(\theta_A + \theta_\mu)\tan \beta,
\label{twoloop}
\ea
where $m_A$ is the mass of the pseudoscalar Higgs boson, that we took to be
smaller than $M_{\rm SUSY}$, $Y_t\simeq 1$ is the top quark Yukawa coupling in the SM, and 
$\kappa_e \simeq 0.6\times 10^{-25}$cm. For very large values of $\tan\beta$ additional 
contributions from sbottom and stau loops, which are enhanced by higher powers of $\tan\beta$,
also have to be taken into account  \cite{dlopr,twoloop}. 
 
Finally, the second, and in some sense more profound change is that at large 
$\tan\beta$, the observable EDMs of neutrons and heavy atoms receive significant contributions 
not only from the EDMs of the constituent particles, e.g. $d_e$ and $d_q$, but also from $CP$-odd  
four-fermion operators \cite{LP}. The relevant Higgs-exchange diagram is shown in Figure~9. 
The $CP$ violation in the Higgs-fermion vertex originates from the $CP$-odd correction to 
the fermion mass operator in Figure~7. These diagrams, since they are induced by 
Higgs exchange, receive an even more significant enhancement by $(\tan \beta)^3$. 
In the same approximation as before, the value of the thallium EDM 
induced by this Higgs-exhange mechanism, and normalized to the current experimental limit, 
is given by
\ba
\frac{d_{\rm Tl}}{[d_{\rm Tl}]_{{\rm{ exp}}}}\simeq \frac{\tan ^{3}\beta }
{330 } \left(\fr{100{\rm GeV}}{m_A} \right)^2\Big[ \sin \th _{\mu }+0.04\sin
(\theta_{\mu }+\theta_A)\Big].
\label{Tlsimple}
\ea 
Notice that this result {\em does not} scale to zero as $M_{\rm SUSY}\to \infty$. Although  just 
an $O(10^{-3}-10^{-2})$ correction for $\tan\beta\sim O(1)$, these Higgs--exchange contributions
become very large for $\tan\beta\sim O(50)$ \cite{LP,pilaftsis02,dlopr} (see also \cite{barr}).

\begin{figure}[t]
 \centerline{%
   \includegraphics[width=12cm]{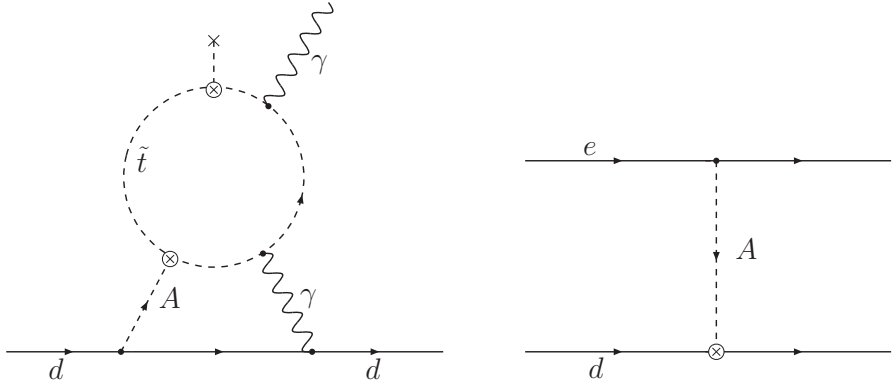}
         }
\vspace{0.1in}
 \caption{\small Additional corrections to the EDMs. On the left two-loop Barr-Zee type graphs
mediated by a stop-loop and a pseudoscalar Higgs, while on the right we have a Higgs-mediated 
electron-quark interaction $C_{de}$ with $CP$ violation at the Higgs-quark vertex. There 
is a second diagram with $CP$-violation at the Higgs-electron vertex mediated by $H$.}
\end{figure}

\subsubsection{The SUSY $CP$ problem}
 
Figure 8 exemplifies the so-called SUSY $CP$ problem: either the $CP$-violating 
phases are small, or the scale of the soft-breaking masses is significantly larger than 1TeV, or  
schematically,
\ba
\delta_{CP} \times \left(\fr{1{\rm TeV}}{M_{\rm SUSY}}\right)^2 < 1.
\ea
The need to provide a plausible explanation to the SUSY $CP$ problem has spawned a sizable 
literature, and the following modifications to the SUSY spectrum have been discussed.
\begin{itemize}
\item {\it Heavy superpartners}. If the masses of the supersymmetric partners exhibit 
certain hierarchy patterns the SUSY $CP$ problem can be alleviated. One of the more
actively discussed possibilities is an inverted hierarchy among the slepton and 
squark masses, i.e. with the squarks of the first two generations being much heavier than 
the stops, sbottoms and staus, ie. $({\bf M}^2_S)_{ij} \gg ({\bf M}^2_S)_{i3},~({\bf M}^2_S)_{33}$,
where $i,j=1,2$ is the generation index \cite{effectivesusy}. It is preferable to have 
masses of the third generation sfermions under the TeV scale because they enter into 
radiative corrections to the Higgs potential, and making them too heavy would re-introduce 
the fine-tuning of the Higgs mass whose resolution was one of the primary motivations for weak-scale
SUSY. Such a framework suppresses the one-loop EDMs which become immeasurably small if the 
scale of the $u$ and $d$ squarks is pushed all the way to $\sim 50$TeV, as suggested by the absence 
of SUSY contributions in $\Delta m_{K(B)}$. This does not mean, however, that the EDMs in such models become 
comparable to $d_i(\delta_{\rm KM})$. Indeed the two-loop contributions to $d_i$ and $w$ 
involving the third generation sfermions are not small in this framework, and indeed 
are at (or sometimes above) the level of current experimental sensitivity. Also, this means of
suppressing the EDMs would not necessarily work in the large $\tan\beta$ regime where 
Higgs exchange may induce a large value for $C_S$ that is not as sensitive to $M_{\rm SUSY}$ as 
the EDM operators. We note that future improvements in experimental precision will 
allow a stringent probe of such scenarios.

\item {\it Small phases}.
A rather obvious possibility for suppressing EDMs is the assumption of an exact (or approximate) $CP$
symmetry of the soft-breaking sector. This is essentially a ``model-building'' option, 
and various ways of avoiding the SUSY flavour and $CP$ problem in this way 
have been suggested in the past fifteen 
years \cite{GaugeMed,AnomMed,OtherMad}. The idea of using low-energy supersymmetry breaking 
looks especially appealing, as it can also help in constructing an axion-less 
solution to the strong $CP$ problem \cite{HilSm}. If the $CP$-odd phases in the soft-breaking sector are 
exactly zero and the conditions (\ref{prop}) are imposed {\it exactly} at the unification scale
as a constraint on the high scale model, we can ask about the scale of EDMs induced by SUSY diagrams  
due purely to $\delta_{\rm KM}$. Since such an MSSM framework would possess the same 
flavour properties as the SM, one expects proportionality to the same $CP$-odd invariant 
combination of mixing angles, namely $J_{\rm CP}$, and suppression by differences of 
Yukawa couplings \cite{Hall}. Then it is easy to understand that the
superpartner contributions to the down quark (C)EDM will necessarily be suppressed by 
the equivalent of $\delta_{\rm CP} \sim JY_c^2 \sim 10^{-9}$, which is again six to seven 
orders of magnitude below current experimental capabilities, and thus not significantly 
larger than  the EDMs induced in the SM \cite{notlarge}. 

\item {\it Accidental cancellations}. Another possibility entertained in recent years \cite{IN,cancel}
is the partial or complete cancellation between the contributions of several $CP$-odd 
sources to physical observables, thus allowing for $\delta_{\rm CP}\sim O(1)$ 
with $M_{\rm SUSY}<1$~TeV. Since the number of potential $CP$-odd phases is large, and the
superpartner mass spectrum is clearly unknown,
one cannot exclude this possibility in principle. However, 
as we illustrated in Figure 8, $d_n$, $d_{\rm Tl}$ and $d_{\rm Hg}$ depend on different 
combinations of phases, and the possibility of such a cancellation looks improbable. 
A more thorough exploration of the MSSM parameter space in search of acceptable 
solutions that pass the EDM constraints was performed in \cite{Barger:2001nu,Abel:2001mc},
and in the absence of additional parameter tuning did not identify any significant 
regions of cancellation.

\item {\it No electroweak scale supersymmetry}. Of course, there is always the possibility that other 
mechanisms (or no easily identifiable mechanism at all) lie behind the gauge hierarchy problem
and the SM is a good effective theory valid up to energy scales {\em much larger} than 1 TeV. 
In this case there is no SUSY $CP$ problem by definition. One of the recently suggested 
scenarios \cite{split} exploits the possibility of a large number of electroweak vacua to invoke 
anthropic reasoning for selecting the ``right'' vacuum, thus side-stepping naturalness arguments
for expecting new physics at the weak scale. Ref.~\cite{split} assumes that 
all the scalar superpartners are very heavy, but leaves gauginos and 
Higgsinos under a TeV, in order to preserve gauge-coupling unification and a dark matter candidate. 
This eliminates the one-loop induced EDMs, but leaves room for two-loop contributions 
\cite{twoloop,quickguys} generated by chargino loops via a diagram similar to that shown
in Figure~9 with $A$ replaced by the light Higgs. This scenario can also be probed with
the predicted sensitivity of future EDM experiments.
 
\end{itemize}

\subsection{SUSY EDMs from flavour physics}

EDMs can also serve as a sensitive probe of non-minimal flavour physics. 
Indeed, the assumptions of proportionality and universality in the soft-breaking sector
(\ref{prop}) at a given high-energy scale are highly idealized, and are not expected 
to hold with arbitrary precision. In this subsection, we would like to show that EDMs are sensitive to 
flavour-changing terms in the soft-breaking sector, and provide 
significant constraints on SUSY models with non-minimal flavour structure. 

For concreteness, let us assume that (\ref{prop}) holds 
approximately, and the perturbations are small. Around the electroweak 
scale, and in a basis with diagonal quark mass matrices, 
the soft-breaking mass matrices can be approximated as 
\ba
{\bf M}^2_S = {\rm diag}( m^2_{S11},m_{S22}^2,m^2_{S33}) + \delta M^2_{Sij},
\ea
where, as before, $S$ labels the different squarks and sleptons, and $i\neq j$.  
Using this approximation, we can calculate the contributions to the 
relevant observables using $\delta M^2_{Sij}$ as a perturbastion via insertions along the 
squark line, as in Figure~10. 

\begin{figure}[t]
 \centerline{%
   \includegraphics[width=7cm]{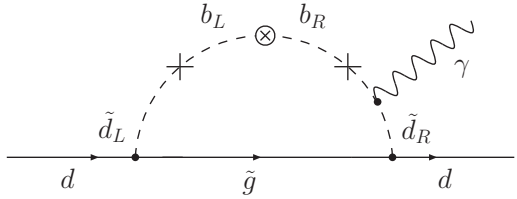}
         }
\vspace{0.1in}
 \caption{\small Contribution of flavour changing processes 
to the $d$-quark EDM. The middle insertion on the sfermion line corresponds to $LR$ mixing 
proportional to $m_b$; the insertions on the left and on the right correspond to flavor transitions 
in the $LL$ and $RR$ squark mass sector. }
\end{figure}

Calculating the gluino one-loop diagram
in the approximation of equal SUSY mass scales, $({\bf M}^2_S)_{ii} = M_i^2 = |\mu|^2=M_{\rm SUSY}^2$, 
we arrive at the following result
for the $d$ quark EDM, and the imaginary correction to the $d$ quark mass, 
\ba
\label{131}
{\rm Im}\, m_d = -\delta^d_{131}\times m_b ~\fr{\alpha_s \tan\beta}{18\pi }\;\;\;\; \nonumber\\
d_d = \delta^d_{131} \times e_d m_b~\fr{\alpha_s \tan\beta }{45\pi M_{\rm SUSY}^2} ,
\label{flavourresults}
\ea
where $\delta^d_{131}$ denotes the following $CP$-odd dimensionless combination,
\ba
\delta^d_{131} = \fr{{\rm Im}( \delta M_{Q13}^2 e^{i\theta_\mu} \delta M_{D31}^2)}{M_{\rm SUSY}^4}.
\ea
In (\ref{131}), for simplicity, we neglected the contributions from the $A$ parameters, 
and retained only the mixing coefficients between the first and the third generations. 
There are two important points about Eq.~(\ref{131}) that we should emphasize here: $\delta_{131}$ 
can be non-zero even if $\th_\mu = 0$, and both Im$(m_d)$ and $d_d$ are enhanced 
relative to (\ref{oneloopth}) and (\ref{simple_d}) by the large ratio $(m_b/m_d) \sim 10^3$, which
can compensate the suppression associated with flavour violation. In the case of
$u$ quark operators, this enhancement factor is even larger, $m_t/m_u \sim 10^5$.  

As we have seen in the previous subsection, renormalization of $\bar\th\sim$Im$(m_q)/m_q$ 
can be very large, capable of producing bounds on $\delta^d_{131}$ and $\delta^u_{131}$ at 
the $10^{-9}$ level or better unless $\bar\theta$ is removed via PQ symmetry. In the latter 
case, using (\ref{131}) and similar results in the lepton sector, one obtains the following 
sensitivity of EDMs to the above combination of flavour changing 
transitions on electron, $u$, and $d$ quark lines for $M_{\rm SUSY} = 1$TeV,
\ba
\delta^e_{131}\sim 10^{-4} - 10^{-3};~~\delta^u_{131}\sim 10^{-6}-10^{-5};
   ~~\delta^d_{131}\sim 10^{-4} - 10^{-3}. \label{fl_constr}
\ea
Thus, EDMs independently provide  very stringent constraints on the combined sources of 
flavour- and $CP$-violation in the soft-breaking sector. These constraints 
are complementary to those coming from  $K$ and $B$ meson physics and searches for
lepton flavour violation. Note also that these calculations need to be modified in the 
large $\tan\beta$ regime to include Higgs-mediated contributions \cite{HPT,LP}
which may dominate over the one-loop results for $d_i$ (\ref{flavourresults}). 

It is important to realize that the apparent enhancement of EDMs in (\ref{131}) by the ratios of
heavy to light quark and lepton masses occurred because of the presence of flavour-changing 
terms in {\em both} $LL$ and $RR$ sectors of the squark/slepton mass matrices. Indeed, to make this
point transparent we can write
\ba
 \de_{131}^d = {\rm Arg}[(\de_{13}^d)_{LL}(\de_{33}^d)_{LR}(\de_{31}^d)_{RR}],
\ea
in terms of ``mass insertions'' \cite{mi} $\de^f_{ij} = M^2_{\tilde{f}ij}/m^2_{\tilde{f}}$, 
which, although the distinction is not crucial here, are usually defined in 
a slightly different basis to the one we have been using. $m^2_{\tilde{f}}$ denotes here the average sfermion
mass-squared. The status of the $LL$ and $RR$ insertions is in general rather different, and particularly so 
within the MSSM where the latter are essentially absent.

To see this in more detail, we recall that flavour-changing terms in the $LL$ sector are natural, 
as they are induced by renormalization group evolution of the soft-breaking parameters 
(see {\rm e.g.} \cite{MV}) even if one assumes the conditions (\ref{prop}) at the 
unification scale. Starting from the universal boundary conditions (\ref{prop}) for all scalar masses,
equal to $m_0^2$, and $A$ parameters at some high-energy scale $\Lambda_{UV}$, one can 
obtain an expression for ${\bf M}^2_Q$ at a lower energy scale $\Lambda$, which at one-loop
is given by 
\ba 
{\bf M}_{Q}^2 = m_0^2~ {\bf 1} - \fr{3m_0^2+A^2}{16\pi^2}
\ln\left(\fr{\Lambda_{UV}^2}{\Lambda^2}\right)\left({\bf Y}_u^\dagger{\bf Y}_u + 
{\bf Y}_d^\dagger {\bf Y}_d\right)+\cdots
\label{YY}
\ea 
The ellipsis denotes ``flavour-blind'' contributions and also higher-order terms. 
Depending on the particular model of SUSY breaking $\Lambda_{UV}$ can be anywhere between 
a few tens of TeV and the Planck scale.  
The presence of {\em both} up and down Yukawa matrices in (\ref{YY}) guarantees 
the appearence of flavour-changing contributions in the $LL$ entries of the 
squark mass matrices. At the superpartner threshold, $\Lambda = M_{\rm SUSY}$,  the 
flavour changing terms in the down squark sector will evidently be 
\ba
\delta M^2_{\tilde dij} \simeq -\fr{3m_0^2+A^2}{16\pi^2}
\ln\left(\fr{\Lambda_{UV}^2}{M_{\rm SUSY}^2}\right)Y_t^2 V_{ti}^* V_{tj}, 
\ea
where $V$ is the CKM matrix, and $Y_t$ is the top quark Yukawa coupling. 
If the scale $\Lambda_{UV}$ is very high, i.e. comparable to the Planck
or GUT scale, the logarithm is large and can entirely offset the loop factor. 
Therefore, the natural size of the 13 entry in the down squark $LL$ sector 
is $\sim M_{\rm SUSY}^2 V_{td}\simeq 0.01 M_{\rm SUSY}^2$.

The situation in the $RR$ sector is completely different. There the absence 
of any ${\bf Y}_u$-dependence in the RG equations for ${\bf M}_D^2$ forbids the generation of
substantial flavour-changing transitions, unless the MSSM spectrum is modified 
above certain energies so that the RG equations for the right-handed squark 
masses aquire flavour dependence. 

A number of SUSY scenarios have been proposed which describe plausible patterns of small 
deviations from (\ref{prop}), allowing for significant $RR$ contributions.
In models with SO(10) unification and  $\Lambda_{UV}>\Lambda_{\rm GUT}$
the running of the soft-breaking parameters extends {\em above} the 
unification scale, where the RG equations are modified by the presence of new field degrees of freedom. 
For SO(10) GUTs this modification introduces significant
flavour dependence in the $RR$ sector of squark and slepton mass matrices \cite{DH},
even if the restrictions (\ref{prop}) are imposed at the Planck scale. 
The resulting flavour-changing terms for down squarks $\delta  M^2_{Dij}$ are of the same 
order of magnitude as in the $LL$ sector, leading to the prediction  $\delta^d_{131}\sim 10^{-4}$, which is 
right at the border-line of current experimental sensitivity (\ref{fl_constr}), \cite{so10,Zyablyuk}.
Similar effects may be generated by heavy sterile neutrinos. The light neutrino mass 
scale might, via the seesaw mechanism, be 
pointing to the existence of a new energy scale, $M_R \sim Y^2_\nu v^2/(0.1-0.001~{\rm eV})$ 
related to heavy sterile (or ``right-handed'') neutrinos. If $\Lambda_{UV}$ is larger than 
$M_R$, the RG equations for sleptons will be modified above $M_R$ with an effect similar to that above, 
namely a non-trivial flavour dependence will be imprinted on the slepton mass matrices. 
The importance of such an effect will depend on the size of the neutrino Yukawa couplings $Y_\nu$ and, 
with certain Yukawa patterns, an observable or nearly-observable electron EDM might be induced
\cite{nuR}. Of course, if the scale of SUSY breaking is lower than $M_R$ 
(or $\Lambda_{GUT}$) there are no significant consequences for EDMs unless one allows for
other ``diagonal'' phases in this sector.

\section{Conclusions and Future directions }

Recent years have seen significant progress in the experimental tests of $CP$-violation in 
the Standard Model. Experimental verification of direct $CP$ violation in Kaon decay, and in
particular the 
spectacular measurements of $CP$ asymmetries for neutral $B$ meson decays at BaBar and Belle have
provided solid confirmation of the overall validity of the Kobayashi-Maskawa mechanism. 
The current status of $CP$ violation in flavour changing processes is such that (within errors)
it does not necessitate the introduction of any additional $CP$-violating sources. 
At the same time, there is ample (experimental) room for the existence of new 
$CP$-violating physics to which the $K$ and $B$ meson 
data is not sensitive. This concerns, primarily, $CP$ violation in 
flavour-conserving channels. The existence of such new sources is hinted at, albeit indirectly,
by the baryon asymmetry in the Universe. The search for $CP$ violation in 
flavour-conserving channels, and the search for EDMs in particular, should thus remain high on
the priority list for particle physics. The strong suppression of EDMs induced purely by the 
Kobayashi-Maskawa phase, combined with prospects for improving the experimental sensitivity, 
places EDM searches at the forefront in probing $CP$-violating physics beyond the Standard Model.

Moreover, beyond their direct sensitivity,
current (and future) null results for EDM searches also provide very powerful constraints 
on models for new physics. Indeed, as we have discussed, 
the sensitivity for example to $CP$ violation in the soft-breaking sector of SUSY models, allows us 
to probe soft-breaking masses as large as a few TeV. In this indirect sense, EDMs are often 
sensitive to energy scales beyond the reach of future collider experiments, and play a central
role in the full suite of precision tests of the Standard Model. The
scales probed by EDMs and by the constraints on flavour changing neutral currents are
not too dis-similar, and the gap may continue to narrow with future progress in EDM searches. This only heightens
the tension between the observed $CP$-violation in the flavour-changing sector and the lack thereof in
flavour-diagonal channels, of which the strong $CP$ problem is the most manifest example. 
We seem compelled to question whether $CP$ and flavour are as intrinisically
linked in general as they are within the Kobayashi-Maskawa model? This is one aspect of what one might
hope would be answered by a general ``theory of flavour''. EDMs will clearly continue to provide a crucial
probe in tackling this question.

In this concluding section we would like to emphasize some directions on the 
experimental and theoretical front that are likely to bring future progress in 
establishing the nature of flavourless $CP$-violation at and above the electroweak scale.

\bigskip
$\bullet$ {\em Experimental Developments}
\bigskip

There are a number of developments in experimental techniques to search for EDMs 
which promise to narrow the gap between the current limits and the KM background in all
of the EDM classes discussed in  this review.
In particular, novel techniques for storing neutrons in liquid helium, in progress
at LANSCE in Los Alamos \cite{lansce} and under investigation for an experiment at PSI, will help 
to improve the measurement of many fundamantal parameters in neutron physics, including 
$d_n$. On another front, the suggestion that $CP$-violating effects can be significantly enhanced
in exotic nuclei possessing an octopole moment \cite{octopole} drives
several experiments searching for EDMs of isotopes of Ra and Rn \cite{RaRn}. We note that such 
experiments should be pursued primarily because of their potential to discover  
$CP$ violation, as null results will not be as constraining as those from $d_{n}$ or $d_{\rm Hg}$ 
due to large uncertainties in the calculations of nuclear matrix elements. In future measurements
of paramagnetic EDMs, the resulting sensitivity to $d_e$ and $C_S$ will be 
significantly improved via the use of paramagnetic molecules, such as YbF and PbO, that can be polarized
and thus allow a huge enhancement of the applied field \cite{YbF,PbO}. The anticipated precision
will allow probes of the electron EDM down to $10^{-30}e$ cm or below \cite{YbF,PbO}. 
In this regard, we should mention an interesting alternative approach, which involves the measurement
of the tiny electron EDM-induced magnetic flux when an electric field is applied to a 
particular garnet crystal. This project is also already in development at LANSCE \cite{garnet}. 

Perhaps the most interesting proposal of recent years is the new approach to 
measure the EDMs of {\it charged} particles in storage rings. An initial proposal to 
measure the muon EDM at the $10^{-24}e$ cm level \cite{muon} has since evolved into the 
idea to measure the EDMs of ions and nuclei \cite{deuteron}, and in particular the deuteron EDM 
at the $10^{-27}e$ cm level. Although clearly 
not a diamagnetic system, the deuteron EDM can be placed in the same category as it is primarily sensitive 
to the same nuclear scale $CP$-violating source, namely the $CP$-odd pion-nucleon couplings $\bar{g}_{\pi NN}$.
More precisely, one finds \cite{kk,lopr,lt},
\ba
 d_D = (d_n + d_p) - (1.3 \pm 0.3\ {\rm GeV}^{-1}) e \bar{g}_{\pi NN}^{(1)}, \label{dD} 
\ea
where the second term generically dominates, leading to a similar dependence on $CP$-odd sources
as for mercury. However, in comparison to mercury, the deuteron EDM proposal has the significant 
advantage of requiring only straightforward nuclear calculations, since it is a weakly bound state, 
which leads to the relatively precise dependence on $\bar{g}_{\pi NN}^{(1)}$ in (\ref{dD}).
Using the result (\ref{goneres}) for $\bar{g}_{\pi NN}^{(1)}$, and accounting for the subleading 
corrections, one obtains \cite{lopr}
\ba
 d_D \simeq 6^{+11}_{-3} e (\tilde{d}_d - \tilde{d}_d) + {\cal O}(\tilde{d}_d +\tilde{d}_u,d_u,d_d).
\ea
Thus a measurement of $d_D$ at the $\sim 10^{-27}e$ cm level would correspond to a sensitivity 
to light quark CEDMs at the level of $10^{-28}$ cm \cite{lopr}, which is 
at least two orders of magnitude better than the current limits from $d_{\rm Hg}$. 
Above all else, these 
proposals to measure EDMs in storange rings deserve special considerations as they depart from 
the dominant philosophy that EDM measurements demand the study of neutral objects in parallel 
electric and magnetic fields. 

To place this activity in context, we should bear in mind that probably the most important single question 
for particle physics -- the origin of  electroweak symmetry breaking -- will be subjected to serious 
experimental scrutiny with the Large Hadron Collider coming on line within a few years. Besides 
the discovery of the Higgs boson(s), it may provide an answer to the gauge hierarchy problem, and 
indeed uncover a plethora of new particles or resonances above the electroweak scale. EDM experiments,
which might of course discover new physics before the LHC switches on, can subsequently 
play a complementary role in providing constraints on (or signatures of) $CP$-violating couplings
(e.g. in the Higgs sector of the MSSM).
The projected level of sensitivity in coming years will be more than competetive  in this 
regard with collider probes. Moreover, strangely enough, the absence of new physics 
(beyond the Higgs -- or whatever
might play this role) at the TeV scale would not remove motivations for EDM searches. Indeed, as we argued 
in this review, EDMs are sensitive to $CP$ violation at multi-TeV scales, and thus represent one of the
few classes of low-energy precision measurements that are sensitive to such high-energy scales.

Another important experimental direction relevant to $CP$-violating physics is 
the search for axions. As we reviewed, one of the more natural resolutions of the strong $CP$ 
problem predicts the existence of a light pseudo-scalar particle, the axion. The developments 
of recent years in cosmology have lent considerable weight to the presence of a non-baryonic 
cold dark matter component of the energy density in the Universe. Although the popularity of 
supersymmetric models continues to focus attention on the lightest supersymmetric particle (or LSP)
as a natural dark matter candidate at the weak scale, axions with a coupling $f_a^{-1}$ below its astrophysical
bound in fact still  represent a viable alternative, thus providing additional motivation
for the continuation of axion searches.

\bigskip
$\bullet$ {\em Theoretical Developments}
\bigskip

On the theoretical side, beyond questions of the precise generation mechanisms of $CP$-odd sources
in specific new physics models, it is clear that the primary limitation on the full application of the
observational bounds arises through the limited precision of QCD and nuclear calculations. Perhaps the
most afflicted quantity at present is the $CP$-odd pion-nucleon constant, as induced in particular
by the CEDMs of light quarks. As we have discussed, this is a fundamental parameter controlling the 
level of the constraints imposed by diamagnetic atoms, which can currently be calculated only to limited
precision due to large cancellations in the relevant nucleon matrix elements. Another important issue 
concerns the strange quark CEDM contribution both to $\bar g_{\pi NN}$ and the neutron EDM $d_n$,
and whether or not it is underestimated in the leading-order sum-rules analysis \cite{FOPR,hs}. It would clearly
be worthwhile to revisit these aspects. However, it seems likely that significant quantitative progress 
will come only from {\em ab initio} lattice calculations. This is a very challenging task, since 
a viable lattice calculation would necessarily have to respect chiral symmetry both at the level of 
quarks and gluons and also among the observable matrix elements between the hadronic states, since this
is the underlying reason for the suppression of $d_n(\bar\theta)$ by $m_*$ and the partial 
suppression of $d_n(\tilde d_q)$. To that end, it will be important to implement a calculation displaying 
all the required symmetries, and in this sense $d_n(\bar\theta)$ would be a good place to start, 
as many features of the answer, such as the dependence on $m_*$ and on 
$\bar \theta = \theta +{\rm arg}~{\rm det} M_q$, 
are enforced by symmetry allowing for independent checks of the calculation.
On the nuclear side, we noted that recent reanalyses of the Schiff moment indicate 
that various many-body effects,
e.g. polarization, can be significant and thus further progress in this area would assist significantly
in improving the quality of constraints on $\bar g _{\pi NN}$ in different isospin channels. It will
also be important, in guiding future experimental ideas, to clarify the size of 
the enhancement of $CP$ violation in exotic nuclei with octupole deformations.

In conclusion, the limits on flavour-diagonal $CP$-violation produced by the null results 
of existing EDM searches already provide strong constraints on new physics
at and above the electroweak scale. Developments in coming years promise to provide us
with a wealth of new information about the nature of $CP$ violation and TeV-scale physics, 
complementary to studies of electroweak symmetry breaking at colliders and flavour studies with $K$ 
and $B$ mesons.

\bigskip
{\bf Acknowledgements}

We would like to thank J. Archambault, A. Czarnecki, D. Demir, T. Falk, C. Hamzaoui, I. Khriplovich, 
O. Lebedev, K. Olive and R. Roiban 
for collaboration on some of the work described
in this review.  The work of M.P. is 
supported in part by the N.S.E.R.C. of Canada.

\end{document}